%% file: ArXivtemplate.tex
\documentclass[letterpaper,10pt]{article}
\usepackage[T1]{fontenc}
\usepackage{parskip}
\usepackage[font={small}]{caption}
\usepackage[margin=2.5cm]{geometry}
\usepackage{times}
\usepackage[affil-it]{authblk}
\usepackage{listings}

\usepackage{booktabs}
\usepackage{makecell}

\date{}

  {}%
  {}

\usepackage{amsmath,amssymb}
\usepackage{graphicx} 

\usepackage{float}
\usepackage[utf8]{inputenc}
\usepackage[english]{babel}
\usepackage{amsthm}
\usepackage{amsmath,amssymb}

\newtheorem{theorem}{Theorem}[section]

\newtheorem{lem}[theorem]{Lemma}

\usepackage{graphicx} 
\usepackage{subfig}
\usepackage{epsfig}
\usepackage{algorithm}
\usepackage{booktabs}

\usepackage{algpseudocode}
\usepackage{float}
\usepackage[english]{babel}
\usepackage{placeins}

\include{usr_macros}

\newcommand{\bmat}[1]{\begin{bmatrix}#1 \end{bmatrix}}
\newcommand{\R}{\mathbb{R}}

\newcommand{\alphaaa}{W}

\usepackage{amsmath}

\begin{document}


\title{Bayesian Parameter Identification for \\
Jump Markov Linear Systems} 


\author{Mark P. Balenzuela\footnote{Corresponding author: Mark.Balenzuela@uon.edu.au}, Adrian G. Wills, Christopher Renton, and Brett Ninness}

\maketitle
Faculty of Engineering and Built Environment, The University of Newcastle, Callaghan, NSW 2308 Australia




\begin{abstract}                          
This paper presents a Bayesian method for identification of jump
Markov linear system parameters.
A primary motivation is to provide accurate quantification of parameter uncertainty without relying on
asymptotic in data-length arguments. 
To achieve this, the paper details a particle-Gibbs sampling approach that provides samples from the
desired posterior distribution. 
These samples are produced by utilising a modified discrete particle filter and carefully chosen conjugate priors.
\end{abstract}


\clearpage
\section{Introduction}
In many important applications, the underlying system is known to
abruptly switch behaviour in a manner that is temporally random. This
type of system behaviour has been observed across a broad range of
applications including econometrics
\cite{kim1994dynamic,hahn2010markov,so1998stochastic,bianchi2012regime,fruhwirth2001markov,hamilton1989new,kim1999state,di2016gibbs},
telecommunications \cite{logothetis1999expectation}, fault detection
and isolation \cite{hashimoto2001sensor}, and mobile robotics
\cite{mazor1998interacting}, to name just a few.

For the purposes of decision making and control, it is vital to obtain
a model of this switched system behaviour that accurately captures
both the dynamics and switching events. At the same time, it is well
recognised that obtaining such a model based on first principles
considerations is often prohibitive.

This latter difficulty, together with the importance of the problem of
switched system modelling has been recognised by many other authors
who have taken an approach of deriving algorithms to estimate both the
dynamics and the switching times on the basis of observed system
behaviour. The vast majority of methods are directed towards providing
maximum-likelihood or related estimates of the system
\cite{blackmore2007model,gil2009beyond,ashley2014sequential,yildirim2013online,svensson2014identification,ghahramani2000variational},
including our own work \cite{EMJMLSpaperBalenzuela}.  While
maximum-likelihood (and related) estimates have proven to be very
useful, their employment within decision making and control should be
accompanied by a quantification of uncertainty. Otherwise trust or
belief may be apportioned in error, which can lead to devastating
outcomes.

Classical results for providing such error quantifications have
relied on central limit theorems, that are asymptotic in the limit as
the available data length~\cite{lehmann1983theory,lennart1999system,astrom1979maximum,aastrom1971system}. These results are then used in
situations where only finite data lengths are available, which can provide
inaccurate error bounds.
This raises a question around providing error quantification that is
accurate for finite length data sets. A further question concerns the
quantification of performance metrics that rely on these system
models, such as the performance of a model-based control strategy.

Fortunately, issues relating to model uncertainty can be addressed using the
Bayesian framework~\cite{robert2007bayesian}. The rationale of the Bayesian approach
is that is captures the full probabilistic information of the system
conditioned on the available data, even for short data lengths. 
However, the Bayesian approach is not without its
challenges. Except for some very special cases, the Bayesian posterior
distribution is not straightforward to compute 
due to the typically large multi-dimensional integrals that
are required in forming the solution.

A remarkable alternative approach aims instead to construct a random
number generator whose samples are distributed according to the
desired Bayesian posterior distribution.  The utility of this random
number generator is that its samples can be used to approximate, with
arbitrary accuracy, user defined expectation integrals via Monte-Carlo
integration (see e.g. \cite{RobertC:2004}). For example, this can
provide estimates of the conditional mean, which is known to be the
minimum mean-squared estimate.

Constructing such a random number generator has received significant
research attention and among the possible solutions is the so-called
Markov chain approach. As the name suggests, this approach constructs
a Markov chain whose stationary distribution coincides with desired
target distribution, and therefore iterates of the Markov chain are in
fact samples from the desired target.

Ensuring that this Markov chain does indeed converge to the desired
stationary distribution can be achieved using the Metropolis--Hastings
(MH) algorithm (see e.g. \cite{RobertC:2004}). Despite achieving this
remarkable outcome, the MH algorithm approach has a significant
drawback in that it requires a certain user supplied ``proposal
density''. This proposal must be carefully selected so as to deliver
realisations that have a reasonable probability of originating from the target distribution.

Fortunately, this latter problem has received significant research
attention and there are by now many variants of the MH method. 
One alternative, the Gibbs sampler, proceeds by sampling
from conditional distributions in a given sequence with the primary
benefit that each sample is guaranteed to be accepted.
Alternatively, as we will consider in this paper, particle filters may be used to construct samples from the required conditional distribution, this resulting algorithm is aptly referred to as the particle-Gibbs method. 

This particle-Gibbs approach has been previously employed for identifying linear
state-space models \cite{wills2012estimation,carter1994gibbs},
autoregressive (AR) systems
\cite{diebolt1994estimation,fruhwirth2001markov,gerlach2000efficient,albert1993bayes},
change-point models
\cite{chib1998estimation,carter1996markov,giordani2008efficient},
stochastic volatility models \cite{so1998stochastic}, and
stochastically switched systems operating according to drift processes
\cite{hahn2010markov}.
Employing the particle-Gibbs sampler for a switched system encounters
additional challenges. In essence, these stem from the exponential (in
data-length) growth in complexity caused by possibility that the
system switches among a number of possible dynamic models at each
time-step
\cite{generalJMLSpaperBalenzuela,ashley2014sequential,blackmore2007model,gil2009beyond,balakrishnan2004inference,ghahramani2000variational,blom1988interacting,bergman2000markov,barber2006expectation,helmick1995fixed,kim1994dynamic}.

In this work, we focus on applying a particle-Gibbs sampler to an important subclass of switched systems
known as jump Markov linear systems (JMLS). This class is attractive
due to its relative simplicity and versatility in modelling switched
systems.
Previous research effort has considered particle-Gibbs sampling for the JMLS
class, each with various limitations, including univariate models in
\cite{kim1999state,fruhwirth2001fully,fruhwirth2001markov}, limited
dynamics modes in \cite{kim1999state,chang2018state,albert1993bayes},
and constrained system or noise structure in
\cite{carter1996markov,hahn2010markov,giordani2008efficient,gerlach2000efficient,fox2011bayesian}.
We focus on removing these limitations.

To perform particle-Gibbs sampling on this system class two methods need to be provided:
\begin{enumerate}
\item A method for generating smoothed hybrid state trajectories conditioned on the available data.
\item A method of updating the prior chosen for the system parameters using the sampled hybrid state trajectory. 
\end{enumerate}
The latter can employ the MH algorithm, but is attended to by use of conjugate prior solutions in this paper as this approach is computationally advantageous. 
The relevant details for this approach are discussed later.

Previous solutions to the former problem have involved sampling the continuous and discrete latent variables in separate stages of the particle-Gibbs algorithm \cite{shephard1994partial,kim1999state,fruhwirth2001fully,fruhwirth2001markov,carter1996markov,gerlach2000efficient,so1998stochastic,fox2011bayesian}.
These approaches rely on \cite{fruhwirth1994data,fruhwirth2006finite} for sampling the continuous state trajectory, and \cite{diebolt1994estimation,hamilton1989new,chib1996calculating,fruhwirth2006finite} for the sampling the discrete state sequence.
Methods used to sample the discrete sequence include `single move' sampling, where the model used at each time step is sampled separately by increasing the number of steps within the particle-Gibbs iteration, and `multi-move sampling' where the entire discrete trajectory is sampled as a block \cite{fruhwirth2006finite,kim1999state}.
%
However, these approaches have been criticised for poor mixing \cite{whiteley2010efficient}, slowing down the convergence rate of the particle-Gibbs algorithm.

Rao-Blackwellization may be used to improve the efficiency of particle filters. 
Additional improvements often used with Rao-Blackwellization include using a discrete particle filter (DPF) \cite{fearnhead2003line}, which offers more efficient sampling from the discrete state space. 
The algorithm does this by using a metric to determine which components/particles to be kept deterministically, and which particles should undergo a resampling procedure. 
As the particles which are kept deterministically are those with the greatest probability mass, and therefore are not duplicated during resampling, the algorithm presents with fewer duplicated particles, and is therefore more efficient.
The DPF algorithm also promises to be exact as the number of particles reaches the impracticable limit of the total number of components required to represent the probability distribution exactly, as all components will be kept deterministically in this case.

In this paper, we apply the particle-Gibbs sampler to the general JMLS class using a modified version of the DPF.
As detailed later, this modified version is specifically tailored to the JMLS class and attends to some of the shortcomings in the papers which conceived the idea \cite{fearnhead2003line,whiteley2010efficient}.
Unlike previous particle-Gibbs approaches to JMLS identification, we do not assume the system to be univariate \cite{kim1999state,fruhwirth2001fully,fruhwirth2001markov}, support only a small number of models \cite{kim1999state,chang2018state,albert1993bayes}, or constrain the system or noise structure \cite{fox2011bayesian,carter1996markov,hahn2010markov,giordani2008efficient,gerlach2000efficient}.
This in part, is made possible by use of the inverse-Wishart conjugate prior such as used within \cite{wills2012estimation,fox2011bayesian}, opposed to commonly used inverse-Gamma distribution (e.g. see \cite{albert1993bayes,chib1996calculating,kim1999state,fruhwirth2001markov,diebolt1994estimation}) that only describes scalar probability distributions. 

\textbf{{\em The contributions}} of this paper are therefore:
\begin{enumerate}
\item A self contained particle-Gibbs sampler which targets the parameter distribution for 
JMLS systems, \emph{without} restriction on state dimension, model 
or noise structure, or the number of models. 
\item The provided solution uses a modified discrete particle filter for application specifically toward JMLS, which is detailed thoroughly. 
A discussion is also provided, clarifying a point of confusion in the original papers \cite{whiteley2010efficient,fearnhead2003line}.
Additionally, an alternative sampling approach is used when compared to \cite{whiteley2010efficient}, which is simpler and unbiased.
\end{enumerate}
%
%
\section{Problem Formulation}\label{sec:problem-formulation}
In this paper, we address the modelling of systems which can jump
between a finite number of linear dynamic system modes using a 
discrete-time jump Markov linear system description, which can be expressed as
\begin{subequations}
\label{eq:JMLSdef1}
\begin{align}
  {\bmat{y_k \\ x_{k+1}}}&= \underbrace{\bmat{\mathbf{C}_{z_k} &                             \mathbf{D}_{z_k}
  \\ \mathbf{A}_{z_k} & \mathbf{B}_{z_k}}}_{\triangleq \boldsymbol{\Gamma}_{z_k}}
                        {\bmat{x_k  \\ u_k}}
  + \underbrace{\bmat{e_k\\ v_k}}_{w_k},
\end{align}
where $x_k \in \mathbb{R}^{n_x}$ is the system state,
$y_k \in \R^{n_y}$ is the system output, $u_k \in \R^{n_u}$
is the system input, $z_k \in \{ 1,\dots ,m\}$ is a discrete random
variable called the \emph{model index} that indicates the active model, and the
noise terms $v_k$ and $e_k$ originate from the Gaussian
white noise process
\begin{align}
  \label{eq:autosam:3}
  w_k 
  &\sim \mathcal{N}\left(0\ ,\
    \boldsymbol{\Pi}_{z_k} \right ),
    \quad
    \boldsymbol{\Pi}_{z_k} = 
    \begin{bmatrix}
      \mathbf{R}_{z_k}
      &\mathbf{S}^T_{z_k} \\
      \mathbf{S}_{z_k} &
      \mathbf{Q}_{z_k}\end{bmatrix},
\end{align}
\end{subequations}
where
\begin{align}
  \label{eq:14}
\mathcal{N}(x|\mu, \mathbf{P}) = \det(2 \pi
\mathbf{P})^{-\frac{1}{2}} e^{-\frac{1}{2}(x-\mu)^T
  \mathbf{P}^{-1}(x-\mu)}  
\end{align}
denotes a multivariate Normal distribution.

The system matrices and noise covariances are allowed to randomly jump
or switch values as a function of the model index $z_k$. A switch
event is captured by allowing $z_k$ to transition to $z_{k+1}$
stochastically with the probability of transitioning from the
$j_\text{th}$ model at time-index $k$ to the $i_\text{th}$ model at
time-index $k+1$ given by
\begin{align}
  \label{eq:JMLSdef2}
  \mathbb{P}(z_{k+1} = i | z_k = j) &= \mathbf{T}_{i,j}, \quad \sum_{i=1}^{m} \mathbf{T}_{i,j} &= 1 \quad  \forall j.
\end{align}
The set of model parameters that fully describe the above
JMLS class can be conveniently collected into a parameter object
$\theta$, defined as
\begin{align}
  \label{eq:autosam:4}
    \theta \triangleq \bigl \{ \mathbf{T}, \{ \boldsymbol{\Gamma}_i, \boldsymbol{\Pi}_i  \}_{i=1}^m \bigr \}.
\end{align}
\textbf{Problem:} {\em Presented with known state dimension $n_x$ and known
number of system modes $m >0$, a data record of $N$ outputs and inputs
\begin{align*}
  \mathbf{y} \triangleq y_{1:N} &= \{y_1,\ldots,y_N \}, \quad 
  \mathbf{u} \triangleq u_{1:N} = \{u_1,\ldots,u_N\},
\end{align*}
respectively, the primary focus of this paper is provide a random
number generator that produces samples $\theta^\ell$ from the target
distribution
\begin{align}
  \label{eq:1}
  \theta^\ell \sim p(\theta \mid y_{1:N}).
\end{align}
}
The following section details how this distribution can be targeted using a Markov chain Monte Carlo technique called particle-Gibbs sampling.
\section{Markov Chain Monte-Carlo Approach}
\label{sec:MCMCapproach}
The essential idea underpinning so-called Markov chain Monte-Carlo (MCMC)
methods is that they are focussed on computing expectation integrals
of the form
\begin{align}
  \label{eq:2}
  I = \int f(\theta)\, p(\theta \mid \mathbf{y}) \text{d} \theta,
\end{align}
where the function $f(\cdot)$ may be quite general. For example, $f$
could be as simple as an indicator function for $\theta$ belonging to
a given set, or more complex functions such as the gain or phase margins
of a $\theta$ dependent control design.

Solving \eqref{eq:2} is generally intractable in closed-form, and
classical quadrature methods are limited to small dimensions
only. An alternative computational approach relies on a law of large
numbers (LLN) to provide estimates of \eqref{eq:2} via so-called Monte-Carlo
integration where samples from $p(\theta \mid \mathbf{y})$ are used to
compute a sample average of $f(\cdot)$ according to
\begin{align}
  \label{eq:3}
  \widehat{I}_M = \frac{1}{M} \sum_{\ell=1}^M f(\theta^\ell), \qquad
  \theta^\ell \sim p(\theta \mid \mathbf{y}).
\end{align}
Under rather general assumptions on $p(\theta \mid \mathbf{y})$ and
$f(\cdot)$, it can be shown that 
\begin{align}
  \label{eq:4}
  \widehat{I}_M \overset{\text{a.s.}}{\to} I \quad \text{as} \quad  M \to \infty,
\end{align}
where a.s. denotes almost sure convergence. Importantly, the rate of
convergence of $\widehat{I}_M$ to $I$ is maximised when the samples
$\theta^\ell$ are uncorrelated~\cite{ninness2000strong}.
This raises the question of how to generate samples from $p(\theta
\mid \mathbf{y})$ that have minimal correlation. 

A remarkably effective approach aimed at solving this
problem is to construct a Markov chain whose stationary distribution
coincides with the target $p(\theta \mid \mathbf{y})$. Equally
remarkable is that such a Markov chain can be constructed in a
straightforward manner using the Metropolis--Hastings (MH)
algorithm~\cite{metropolis1953equation,hastings1970monte}. 

In this work, we will use a variant of the MH algorithm known as the
particle-Gibbs sampler. In essence, the main idea of the particle-Gibbs sampler is to
reduce a high-dimensional sampling problem into a sequence of
low-dimensional sub-problems. The primary aim is to design these
sub-problems so that sampling is relatively straightforward.

In detailing this approach for JMLS identification, it is important to
extend our target distribution to the joint parameter and state
posterior
\begin{align}
  \label{eq:5}
  p(\theta, {\xi}_{1:N+1} \mid \mathbf{y})
\end{align}
where ${\xi}_{1:N+1} $ is a hybrid continuous-discrete trajectory, i.e.,
\begin{subequations}
\begin{align}
\xi_{1:N+1} &= \{ \xi_1 , \dots, \xi_{N+1}\}, \\
\xi_k &= \{ x_k,z_k\}.
\end{align} 
\end{subequations}
The particle-Gibbs algorithm will produce a Markov chain, whose iterates
\begin{align}
  \label{eq:7}
  \{\theta^1, {\xi}^1_{1:N+1}\}, \{\theta^2, {\xi}^2_{1:N+1}\}, \ldots, \{\theta^\ell, {\xi}^\ell_{1:N+1}\},
\end{align}
are samples from the target $p(\theta, {\xi}_{1:N+1} |
\mathbf{y})$. Furthermore, samples from the desired marginal posterior
$p(\theta | \mathbf{y})$ are obtained by simply ignoring the
  ${\xi}_{1:N+1}$ component of the joint samples.

  There are many different ways to design a particle-Gibbs sampler to provide
  the desired Markov chain in \eqref{eq:7}, but possibly the most
  direct choice is to firstly sample the state trajectory
  $\xi_{1:N+1}$ based on the assumption of an available $\theta$
  value. Then in a second step, this is reversed and a new sample of
  $\theta$ is generated based on the sampled state trajectory
  $\xi_{1:N+1}$. This process is then repeated in order to provide the
  desired Markov chain. This procedure can be summarised by the
  following steps, which are indexed by the integer $\ell$ to clarify
  the iterative nature of this approach:
\begin{enumerate}
\item Given $\theta^\ell$, sample a hybrid trajectory according to
\begin{align}
\xi_{1:N+1}^\ell \sim p(\xi_{1:N+1} | \theta^\ell,y_{1:N}).
\end{align}
\item Then, given $\xi_{1:N+1}^\ell$ sample a new $\theta^{\ell+1}$
\begin{align}
\theta^{\ell+1} \sim p(\theta|\xi_{1:N+1}^{\ell},y_{1:N}).
\end{align}
\end{enumerate}
The key to success for the above procedure is tied to the relative
ease of generating samples in each step. Unfortunately, for the JMLS
class it is not tractable to construct
$p(\xi_{1:N+1} | \theta^\ell,y_{1:N})$ exactly, as this distribution
requires an exponentially increasing number of terms as the data
length $N$ increases.

It is tempting to consider the impact of simply replacing the exact
distribution $p(\xi_{1:N+1} | \theta^\ell,y_{1:N})$ with a tractable
approximation instead. While this may seem appealing from a practical
perspective, it in general forfeits guarantees that samples are from the distribution $p(\xi_{1:N+1} | \theta^\ell,y_{1:N})$, and therefore this implies that
the resulting particle-Gibbs sampler does not produce samples from the desired target.




A remarkable result from \cite{andrieu2010particle}
shows that such a particle-Gibbs sampler can be constructed by using sequential
Monte Carlo (SMC) approximations of
$p(\xi_{1:N+1} | \theta^\ell,y_{1:N})$
with guaranteed converge properties if the approximated distribution includes the previously sampled discrete model sequence $z^{\ell-1}_{1:N+1}$.
%
Using this particle approach, a particle filter can sample possible model sequence hypothesis ($z_{1:k}$), indexed by $i$ and $z_k$ to calculate the approximate distribution
  \begin{small}
  \begin{align}
    \label{eq:dpf4654}
    p&(x_k,z_k | y_{1:k}) \approx \sum_{i=1}^{M^f_{k}(z_k)} w^i_{k|k}(z_k) \, \mathcal{N} \left (x_k \big|  \mu^i_{k
                              | k}(z_k),  \mathbf{P}^i_{k | k}(z_k)
                              \right ).
  \end{align}
  \end{small}
As the continuous space has a closed-form solution when conditioned on a model sequence, the particle filter employed for this problem is used to explore the discrete space only.
As such, it is immensely computationally wasteful to directly employ traditional resampling schemes based on the component weights $w^i_{k|k}(z_k)$ with the indices $i$ and $z_k$ collapsed.
Doing so will likely result in duplicate components and repeated calculations, leaving other model hypothesis that could have been considered to remain unexplored.

This motivates the use of the discrete particle filter, which is employed in the proposed scheme to prevent this phenomenon.
As specific details are required to implement a working DPF, it is introduced thoroughly in the following section.
This section begins by relating the approximate distribution \eqref{eq:dpf4654} produced by the DPF, to the required samples $\xi_{1:N+1}^\ell$.

\subsection{Sampling the latent variables}
\label{sec:getlatentvars}

In the proposed solution, the latent variables $\{x^\ell_{1:N+1},z^\ell_{1:N+1} \}$ 
are sampled from an approximated distribution of $p(\xi_{1:N+1} | \theta^\ell,y_{1:N})$, with the requirement that this distribution considers the model sequence $z^{\ell-1}_{1:N+1}$ 
\cite{andrieu2010particle}.
Complying with this requirement is very straight-forward, as any JMLS forward filter with a pruning/resampling scheme can be modified to always produce hybrid mixtures with the component related to $z^{\ell-1}_{1:N+1}$.
Using the produced approximated forward distributions \eqref{eq:dpf4654}, the samples $\{x^\ell_{1:N+1},z^\ell_{1:N+1}\}$ can be generated by first sampling from the prediction distribution $\xi^\ell_{N+1} \sim p(\xi_{N+1}|\theta^\ell,y_{1:N})$, followed by conditioning the filtered distribution $p(\xi_k|\theta^\ell,y_{1:k})$ on the latest sample, and sampling from the resulting distribution. These latter steps can be completed for $k=N,\dots,1$, and can be shown to generate samples from the correct target distribution using the equations 
\begin{small}
\begin{subequations}
\begin{align}
\label{eq:conditionoftraj432}
p&(\xi_{1:N+1}|\theta^\ell,y_{1:N}) \nonumber = p(\xi_{N+1}|\theta^\ell,y_{1:N}) \prod_{k=1}^N p(\xi_{k}|\xi_{k+1:N+1},\theta^\ell,y_{1:N}), \\
p&(\xi_{k}|\xi_{k+1:N+1}^\ell,\theta^\ell,y_{1:N}) = \frac{p(\xi_{k+1}^\ell|\xi_{k},\theta^\ell,y_{1:k})  p(\xi_k|\theta^\ell,y_{1:k})}{p(\xi^\ell_{k+1}|\theta^\ell,y_{1:k})} \propto p(\xi_{k+1}^\ell|\xi_{k},\theta^\ell,y_{1:k})  p(\xi_k|\theta^\ell,y_{1:k}).
\end{align}
\end{subequations}
\end{small}
Care must be taken when choosing the component reduction scheme required to compute $p(\xi_k|\theta^\ell,y_{1:k})$ with a practical computational cost.
A merging-based scheme is strictly prohibited as this voids the convergence guarantees of the particle-Gibbs algorithm.
Additionally, traditional resampling schemes for particles exploring discrete spaces are highly inefficient, as often multiple particles remaining after resampling are clones of each other.
This not only is computationally wasteful, as the same calculations are performed multiple times, but also doesn't explore the discrete space effectively.

This motivates the use of the DPF, a particle filter designed to explore a discrete space.
%
In describing the main ideas behind the DPF it is convenient to define a new set of weights for each time instant, denoted as $W_k$, that are simply the collection of all the weights $w_{k|k}^i(z_k)$ from \eqref{eq:dpf4654} for all values of $i$ and $z_k$. That is
\begin{small}
\begin{align}
 W_k = \{w_{k|k}^1(1),&\dots, w_{k|k}^{M_f(1)}(1), w_{k|k}^1(2), \dots, w_{k|k}^{M_f(2)}(2),  \dots, w_{k|k}^1(m_k),\dots, w_{k|k}^{M_f(m_k)}(m_k) \}.
\end{align}
\end{small}
We can then index this array of weights using the notation $W_k^j$ to indicate the $j^{\text{th}}$ entry.

To reduce the growing number of hypothesises encountered, the DPF uses a sampling scheme which is partially deterministic, where the number of components are reduced but few (if any) clones will exist in the reduced mixture.
Essentially, when tasked to reduce a mixture of $n$ components with respective weights $W_k^j$, which satisfies
$\sum_{j=1}^n \alphaaa_k^j=1$, 
to a mixture with $M$ components, this scheme operates by considering the $c$ corresponding to the unique solution of  
\begin{align}
\label{eq:dpfproblem34}
M = \sum_{j=0}^n \min (c_k \alphaaa_k^j,1).
\end{align}
%
The component reduction algorithm then dictates that components with a weight satisfying
\begin{align}
W_k^j \geq \frac{1}{c_k}
\end{align}
will be kept deterministically, leaving the remainder of the mixture to be undergo resampling.

Interestingly enough, the exact value of $c_k$ doesn't need to be computed, and instead, the inclusion of each of the components can be tested, as detailed in Appendix C of \cite{fearnhead2003line}.
In providing this solution, a second optimisation problem was conceived which only generates values of $c_k$ at the boundaries of the inclusion of each component.
Unfortunately these values of $c$ are in general not common, and confusion mounts as resampled components are allocated a weight of $1/c$, which provably should refer to the value generated from the second optimisation problem.
The $c$'s have been used interchangeably in the paper and has led to dependant literature (e.g. \cite{whiteley2010efficient}) skipping over the second optimisation problem entirely and mistakenly using the $c_k$ from the first optimisation problem. 

To make matters worse, the following issues exist in the provided solution \cite{fearnhead2003line} to the second optimisation problem: 
\begin{enumerate}
\item The solution proposed does not handle the case when all components should be resampled, e.g., if reduction needs to be completed on components with equal weight. 
\item A strict inequality has been used mistakenly to form the sets which are used to compute the terms $A_\kappa$, $B_\kappa$ and $c$ in the paper, misclassifying an equality that will be encountered at every boundary test.
\end{enumerate}

Furthermore, the literature of \cite{fearnhead2003line,whiteley2010efficient} mistakes systematic sampling for stratified sampling, and the sampling algorithm presented in \cite{fearnhead2003line} also contains an error, preventing duplicated components.
These components statistically should either be allowed or instead, the component should be issued a modified weight when sampled multiple times.

The proposed Algorithm builds on the work of \cite{fearnhead2003line,whiteley2010efficient} to provide a simpler DPF algorithm that is free of confusion about a $c$ term, which is capable of handling the case when  all components should undergo resampling.
Like \cite{whiteley2010efficient}, these sequences are produced using a particle-Gibbs algorithm, and the previous discrete state sequence is evaluated in the filter to ensure a guarantee of convergence. 
In the proposed scheme, this component is kept deterministically, which avoids the overly complex conditional sampling scheme from \cite{whiteley2010efficient} that provably places biases upon component number and results in some components never being chosen if the algorithm was rerun an infinite number of times.
To add further distinction to \cite{whiteley2010efficient}, a two-filter smoothing approach is not used in the proposed solution, and instead, the direct conditioning on the partially sampled hybrid trajectory described by \eqref{eq:conditionoftraj432} is used for sampling the remainder of the trajectory. 

As the produced algorithms for sampling a hybrid trajectory are quite long and detailed, they have been placed in Appendix~\ref{sec:hybridsamplingalgos} along with practical notes for implementation. 

With the approach taken for sampling the latent hybrid state trajectory covered, the choice of conjugate priors that govern the parameter distributions are now discussed.
These priors are then updated with the freshly sampled hybrid trajectory $\xi_{1:N+1}^\ell$ to form new parameter distributions $p(\theta|\xi^\ell_{1:N+1},y_{1:N})$, which are subsequently sampled from to produce $\theta^{\ell+1}$.
\subsection{Conditioned parameter distributions}

To enable efficient sampling of the JMLS parameters $\theta$, the parameters were assumed to be distributed according to conjugate priors which allow for $p(\theta|\xi_{1:N+1},y_{1:N})$ to be calculated using closed-form solutions.
These conjugate priors were assumed to be as follows.

Like \cite{diebolt1994estimation,fruhwirth2006finite,chib1996calculating,fruhwirth2001markov,hahn2010markov,giordani2008efficient}, the columns of the model transition matrix $\mathbf{T}$ were assumed to be distributed according to independent Dirichlet distributions,
\begin{align}
\D(\bT|\balpha) = \frac{\Gamma(\sum_{i=1}^m \boldsymbol{\alpha}_{i,j})}{\prod_{i=1}^K \Gamma (\boldsymbol{\alpha}_{i,j})} \prod_{i=1}^m (\mathbf{T}_{i,j})^{\boldsymbol{\alpha}_{i,j}-1},
\end{align}
where $\balpha$ is a matrix of concentration parameters with elements $\boldsymbol{\alpha}_{i,j} > 0$, and $\Gamma(\cdot)$ is the gamma function.

The covariance matrices were assumed to be distributed according to an Inverse-Wishart distribution
\begin{small}
\begin{align}
\IW(\bPi_z| \bLambda_{z}, \nu_{z}) &=\frac{|\bLambda_{z}|^{\nu_{z}/2}}{2^{(n\nu_{z}/2  )}\Gamma_n(\frac{\nu_{z}}{2})}|\bPi_z|^{-(\nu_{z}+n+1)/2}   \exp \left\{-\frac{1}{2} \Tr \left[\bLambda_{z} \bPi^{-1}(z)\right]\right\},
\end{align}
\end{small}
where $\bPi_z\in \real^{n\times n}$  is the positive definite symmetric matrix argument, $\bLambda_z \in \real^{n\times n}$ is the positive definite symmetric scale matrix, $\nu_z\in \real$ is the degrees of freedom, which must satisfy $\nu_z > n-1$, and $\Gamma_n$ is the multivariate gamma function with dimension $n$.
A higher degree of freedom indicates greater confidence about the mean value of $\frac{\bLambda_z}{\nu_z-n-1}$ if $\nu_z > n+1$.

Finally, the deterministic parameters for a model are assumed to be distributed according to a Matrix-Normal distribution with the form
\begin{small}
\begin{align}
\MN&(\bGamma(z) |{\bM_z},\bPi_z,{\bV}_z)   =\frac{\exp \left\{ -\frac{1}{2}\Tr \left[\bV_z^{-1}(\bGamma(z)-\bM_z)^T\bPi^{-1}(z)(\bGamma(z)-\bM_z)\right]\right\}}{(2\pi)^{np/2}|\bV_z|^{n/2}|\bPi_z|^{p/2}},
\end{align}
\end{small}
with mean $\bM\in \real^{n\times p}$,  positive definite column covariance matrix $\bV \in \real^{p\times p}$, and $\Tr(A)$ denoting the trace of matrix $A$.  
The Matrix Normal is related to the Normal distribution by
\begin{align}
\MN(\bGamma|\bM,\bPi,\bV) &= \N(\vect(\bGamma)|\vect(\bM),\bV \otimes \bPi).
\end{align}
Lemma~\ref{lem:correctedconjucate3332} provides instructions on how these conjugate prior distributions can be conditioned on the sample $\xi^\ell_{1:N+1}$, thus calculating the parameters for the distribution $p(\theta|\xi_{1:N+1},y_{1:N})$. 
\begin{lem}
\label{lem:correctedconjucate3332}
\begin{small}
The distribution which parameters are sampled from can be expressed as
\begin{align}
\label{eq:finalsoln76543}
&p(\{\bGamma_i, \bPi_i \}_{i=1}^m, \bT|x^\ell_{1:N+1},z^\ell_{1:N+1},y_{1:N})  \propto    \D(\bT|\bar{\boldsymbol\alpha})   \left(\prod^m_{i=1}  \MN(\bGamma_i |\bar\bM_i,\bPi_i,\bar\bV_i)\IW(\bPi_i| \bar\bLambda_i, \bar\nu_i)  \right),
\end{align}
\end{small}
where the parameters $\{\nu_i, \bM_i,\bV_i,\bLambda_i,\boldsymbol{\alpha}\}$ define the prior for the $i$-th model, and the corrected parameters $\{\bar\nu_i, \bar{\bM}_i, \bar{\bV}_i,\bar\bLambda_i,\bar{\boldsymbol{\alpha}}\}$ for each $i$ can be calculated using
\begin{subequations} 
\label{eq:finaljoint_dist354876}
\begin{align}
\bar\bLambda_i &\define \bLambda_i + \bar\bPhi_i - \bar\bPsi_i{\bar\bSigma_i}^{-1}\bar\bPsi_i^T ,\\
\bar{\nu}_i &\define \nu_i + N_i,\\
\bar\bM_i &\define \bar\bPsi_i{\bar\bSigma_i}^{-1},\\
\bar\bV_i &\define \bar\bSigma_i^{-1},\\
 \bar{\boldsymbol\alpha} &= \boldsymbol\alpha + \mathbf{u},
\end{align}\end{subequations}
which use the quantities, 
\begin{subequations} 
\begin{align}
\bar\bSigma_i &\define \bSigma_i + \bV_i^{-1},\\
 \bar\bPsi_i &\define \bPsi_i + \bM_i\bV_i^{-1},\\
 \bar\bPhi_i &\define \bPhi_i + \bM_i\bV_i^{-1}\bM_i^T ,
 \end{align}
 \begin{align}
 \bPhi_i &\define \sum_{k\in G_i} \begin{bmatrix} y_k \\ x_{k+1}^\ell\end{bmatrix} \begin{bmatrix} y_k \\ x_{k+1}^\ell\end{bmatrix}^T,\\
 \bPsi_i &\define \sum_{k\in G_i} \begin{bmatrix} y_k \\ x_{k+1}^\ell\end{bmatrix}\begin{bmatrix}x_k^\ell\\u_k\end{bmatrix}^T,\\
 \bSigma_i &\define \sum_{k \in G_i} \begin{bmatrix}x_k^\ell\\u_k\end{bmatrix}\begin{bmatrix}x_k^\ell\\u_k\end{bmatrix}^T,\\
 N_i &= \sum_{k\in G_i}1,\\
\mathbf{u}_{j,i} &= \sum_{k\in G_{j,i}} 1,
\end{align} \end{subequations}
where $G_i$ is the set of time steps $k$ for which $z^\ell_k=i$, and $G_{j,i}$ is the set of time steps $k$ for which $z^\ell_{k+1}=j$ and $z^\ell_k=i$.
Note that $\mathbf{u}_{j,i}$ denotes the element in the $j$-th row and $i$-th column of matrix $\mathbf{u}$.
Additionally, $A^T$ and $A^{-1}$ denote the transpose and inverse of matrix $A$ respectively.
\end{lem}

\subsection{Sampling new parameters}
As the distribution the distribution $p(\theta|\xi^\ell_{1:N+1},y_{1:N})$ can now be calculated from Lemma~\ref{lem:correctedconjucate3332}, we now provide instruction on how $\theta^{\ell+1}$ may be sampled from such distribution.

We begin by sampling $\bPi_i$ from a Inverse-Wishart distribution for each model $i=1,\dots,m$, i.e.
\begin{align}
\bPi^{\ell+1}_i \sim \IW( \bar\bLambda_i, \bar\nu_i) \quad \forall i =1,\dots,m.
\end{align}

With $\bPi^{\ell+1}_i$ sampled for $i=1,\dots,m$, it is then possible to sample $\bGamma^{\ell+1}_i$ for each model $i=1,\dots,m$ using Lemma~\ref{lem:sampleGamma33}.
\begin{lem}
\label{lem:sampleGamma33}
If we let $\bGamma^{\ell+1}_i$ be determined by 
\begin{align}
\bGamma^{\ell+1}_i = \bar{\bM}_i + (({\bPi}^{\ell+1}_i)^{1/2})^T \bH_i \bar\bV_i^{1/2}
\end{align}
where $A^{1/2}$ denotes an upper Cholesky factor of $A$, i.e. $(A^{1/2})^TA^{1/2} = A$, and each element of $\bH_i$ is distributed i.i.d. according to 
\begin{align}
h \sim \N(0,1),
\end{align}
then 
\begin{align}
\label{eq:sampleMNdistreq}
{\bGamma}^{\ell+1}_i \sim \MN(\bar{\bM}_i,{\bPi}^{\ell+1}_i,\bar{\bV}_i). 
\end{align}
\end{lem}
\begin{pf}
See \cite{gupta2018matrix}. 
\end{pf}
Finally sampling the transition matrix $\bT$ can be completed by sampling from $m$ Dirichlet distributions, all parametrised by $\bar{\boldsymbol{\alpha}}$.
This is completed by sampling each element of $\bT$ from a Gamma distribution with shape parameter $\bar{\boldsymbol{\alpha}}_{i,j}$, and scale parameter of $1$, i.e., 
\begin{align}
\tilde{\mathbf{T}}_{i,j}\sim \mathcal{G}(\bar{\boldsymbol{\alpha}}_{i,j},1), 
\end{align}
before normalising over each column,
\begin{align}
\mathbf{T}_{i,j}^{\ell+1} = \frac{\tilde{\mathbf{T}}_{i,j}}{\sum_{i} \tilde{\mathbf{T}}_{i,j}}.
\end{align}
The complete procedure for sampling the parameter set is further outlined in Algorithm~\ref{alg:JMLSgibbsparamsample_overview}.
\begin{small}
\begin{algorithm}
\caption{Sampling the Parameters}
\label{alg:JMLSgibbsparamsample_overview}
\begin{algorithmic}[1]
\State{Sample from the Inverse-Wishart distribution $\bPi^{\ell+1}_i \sim \IW (\bPi_i| \bar\bLambda_i,\bar\nu_i) \quad \forall i=1,\dots m$.}
\State{Using Lemma~\ref{lem:sampleGamma33}, sample from the Matrix-Normal distribution $\bGamma^{\ell+1}_i\sim \MN(\bGamma_i|\bar\bM_i,\bPi^{\ell+1}_i,\bar\bV_i) \quad \forall i=1,\dots m$.}
\State{Sample from the Gamma distribution \mbox{$\tilde{\mathbf{T}}_{i,j} \sim \mathcal{G}(\bar{\boldsymbol{\alpha}}_{i,j},1)$}, with shape parameter $\bar{\boldsymbol{\alpha}}_{i,j}$, and scale parameter of 1 $\quad \forall i=1,\dots m, \forall j=1,\dots m$.}
\State{Set $\mathbf{T}_{i,j}^{\ell+1} = \frac{\tilde{\mathbf{T}}_{i,j}}{\sum_{i} \tilde{\mathbf{T}}_{i,j}}\quad \forall i=1,\dots m, \forall j=1,\dots m$.}
\end{algorithmic}
\end{algorithm}
\end{small}
\subsection{Algorithm Overview}
For clarity, the proposed method is summarised in full by Algorithm~\ref{alg:JMLSgibbs_overview}.
\begin{small}
\begin{algorithm}
\caption{Algorithm overview}
\label{alg:JMLSgibbs_overview}
\begin{algorithmic}[1]
\Require State prior $p(x_0,z_0)$, prior on model parameters defined by $\{\nu_i, \bM_i,\bV_i,\bLambda_i\}_{i=1}^m$, prior on model transitions defined by $\boldsymbol{\alpha}$, initial guess of $\theta^1$, which may be provided using the JMLS EM algorithm \cite{EMJMLSpaperBalenzuela}.
\For{$\ell=1$ to Max iterations}
\State{Calculate the Forward-filter distribution, as per Algorithm~\ref{alg:forward_filter}.}
\State{Sample a latent trajectory $\xi^\ell_{1:N+1}$ using the instruction provided in Algorithm~\ref{alg:backwards_sampler}.}
\State{Calculate the conditional parameter distribution $p(\theta|\xi_{1:N+1}^\ell,y_{1:N})$ using Lemma~\ref{lem:correctedconjucate3332}.}
\State{Sample and save new parameter estimate $\theta^{\ell+1} \sim p(\theta|\xi^\ell_{1:N+1},y_{1:N})$ using Algorithm~\ref{alg:JMLSgibbsparamsample_overview}.}
\EndFor
\State{By the convergence properties of the particle-Gibbs sampler, the samples $\theta^\ell$ are now distributed according to $p(\theta|y_{1:N})$.}
\end{algorithmic}
\end{algorithm}
\end{small}
\section{Simulations}\label{sec:simulations}
In this section we provide two simulations demonstrating the effectiveness of the proposed method.
\subsection{Univariate system}
In this example we use the proposed method on a univariate JMLS single-input single-output (SISO) system, comprising of two models $m=2$. 
This ensures that parameters are scalar and distributions can appear on 2D plots.
The choice to estimate a univariate system  also ensures that certain system matrices ($\bA,\bD,\bR$) are free from a similarity transformation \cite{bako2009identification,vidal2002observability,svensson2014identification,EMJMLSpaperBalenzuela}, as in general, there is not a unique solution for these parameters.

The true system used in this example was parameterised by 
\begin{align}
\label{eq:trueparams34}
&\bA_1=0.4766, \bB_1=-1.207,\bC_1=0.233, \bD_1=-0.8935, \bQ_1=10^{-3},\bR_1=0.0202,\bS_1=0,\nonumber \\
&\bA_2=-0.1721, \bB_2=1.5330,\bC_2=-0.1922, \bD_2=1.7449, \bQ_2=0.0340,\bR_2=0.0439,\bS_2=0,\nonumber \\
& \bT=\begin{bmatrix} 0.7 & 0.5\\0.3 & 0.5 \end{bmatrix}.
\end{align}
Input-output data was generated using these parameters and the input $u_k\sim\N(0,1)$ for $N=2000$ time steps before the proposed method was applied to the dataset.
The proposed method used a resampling step allowing $R=5$ hybrid components per time step, and used 
%
%
%
uninformative priors on the parameters to ensure the PDFs were highly data driven.
The priors chosen were parameterised by
\begin{subequations} 
\begin{align}
\bM_1=\bM_2=\mathbf{0}_{2\times 2},\\
\bV_1=\bV_2 = 13\cdot\mathbf{I}_{2\times 2},\\
\bLambda_1=\bLambda_2=10^{-10}\cdot\mathbf{I}_{2\times 2},\\
\nu_1=\nu_2=2,\\
\boldsymbol{\alpha} = \mathbf{1}_{2\times 2}.
\end{align}
\end{subequations}
Using these priors, the PDFs produced by the proposed method should have good support of the true parameters, and grow certainty about them with increasing size of dataset.
Other alternative algorithms \cite{kim1999state} can potentially operate on this system, but due to their use of inverse-Gamma distributions, cannot operate on the example within subsection~\ref{sec:example2}. Additionally, as a univariate inverse-Wishart distribution is an inverse-Gamma distribution, these algorithms are equivalent for the univariate case.

The initial parameter set used in the particle-Gibbs sampler is allowed to be chosen arbitrarily.
To avoid a lengthy burn-in procedure, the particle-Gibbs algorithm was initialised with values close to those which correspond the maximum likelihood solution.
For a real-world problem this could be provided using the EM algorithm~\cite{EMJMLSpaperBalenzuela}.
%

\begin{figure}
     \subfloat[Probability distribution of consecutive use of the first model.]{%
       \includegraphics[width=0.4\textwidth]{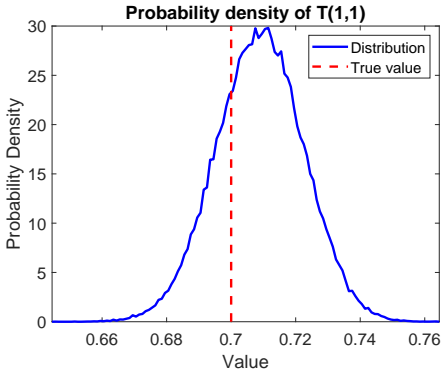}
     }
     \hfill
     \subfloat[Probability distribution of consecutive use of the second model.]{%
       \includegraphics[width=0.4\textwidth]{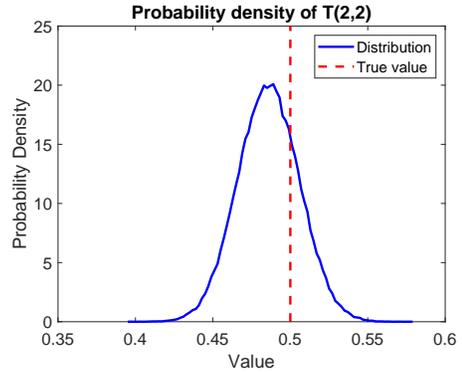}
     }
     \caption{Estimated probability distributions of the transition matrix $\bT$ from Example 1. The distribution from the proposed method is shown in solid blue, whereas the true values are indicated by a red dashed vertical line.}
\label{fig:transitionmatrixprobex1}
\end{figure}

\begin{figure*}
     \subfloat[Probability distribution of $\bA_1$.]{%
       \includegraphics[width=0.45\textwidth]{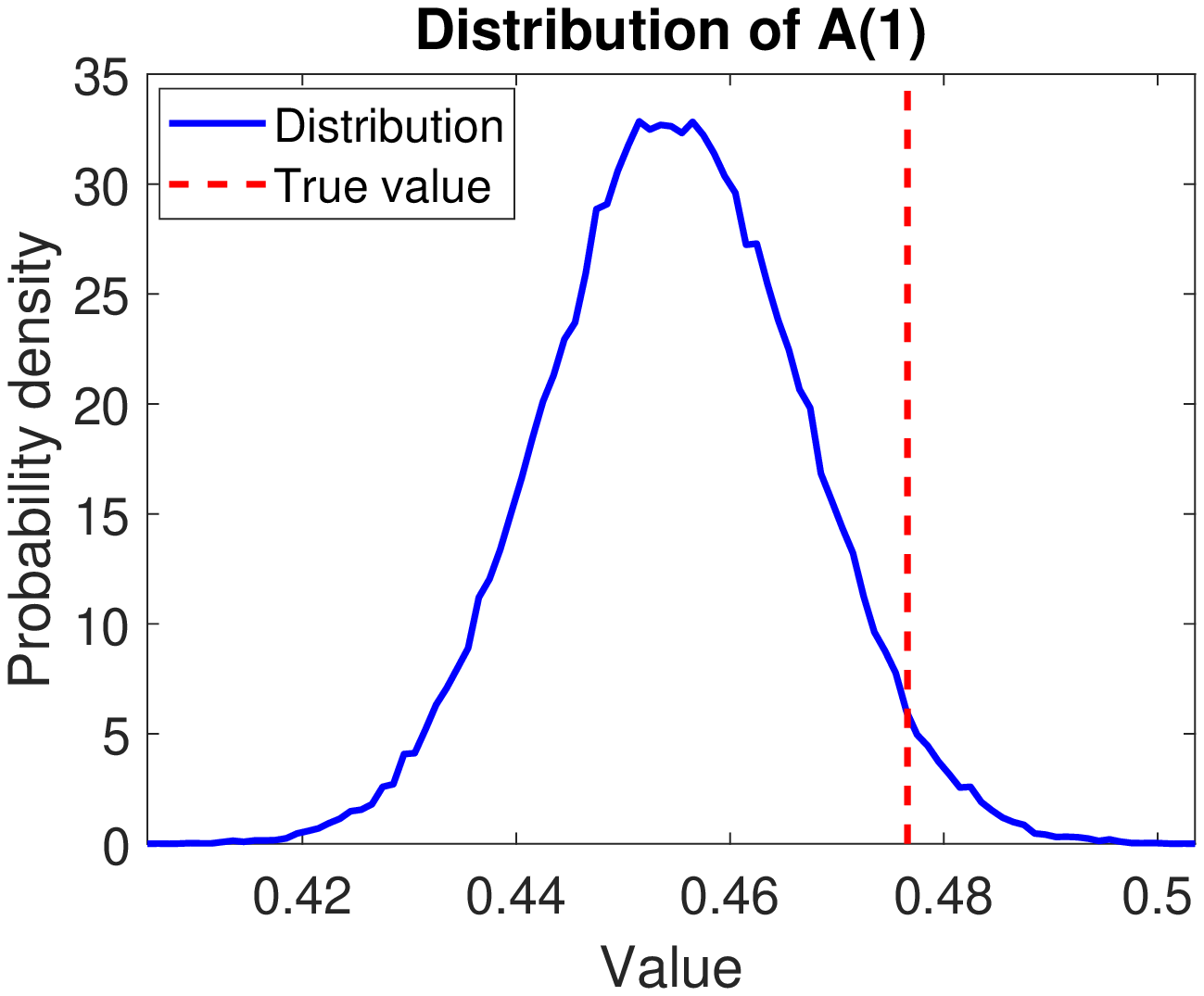}
     }
     \hfill
     \subfloat[Probability distribution of $\bA_2$.]{%
       \includegraphics[width=0.45\textwidth]{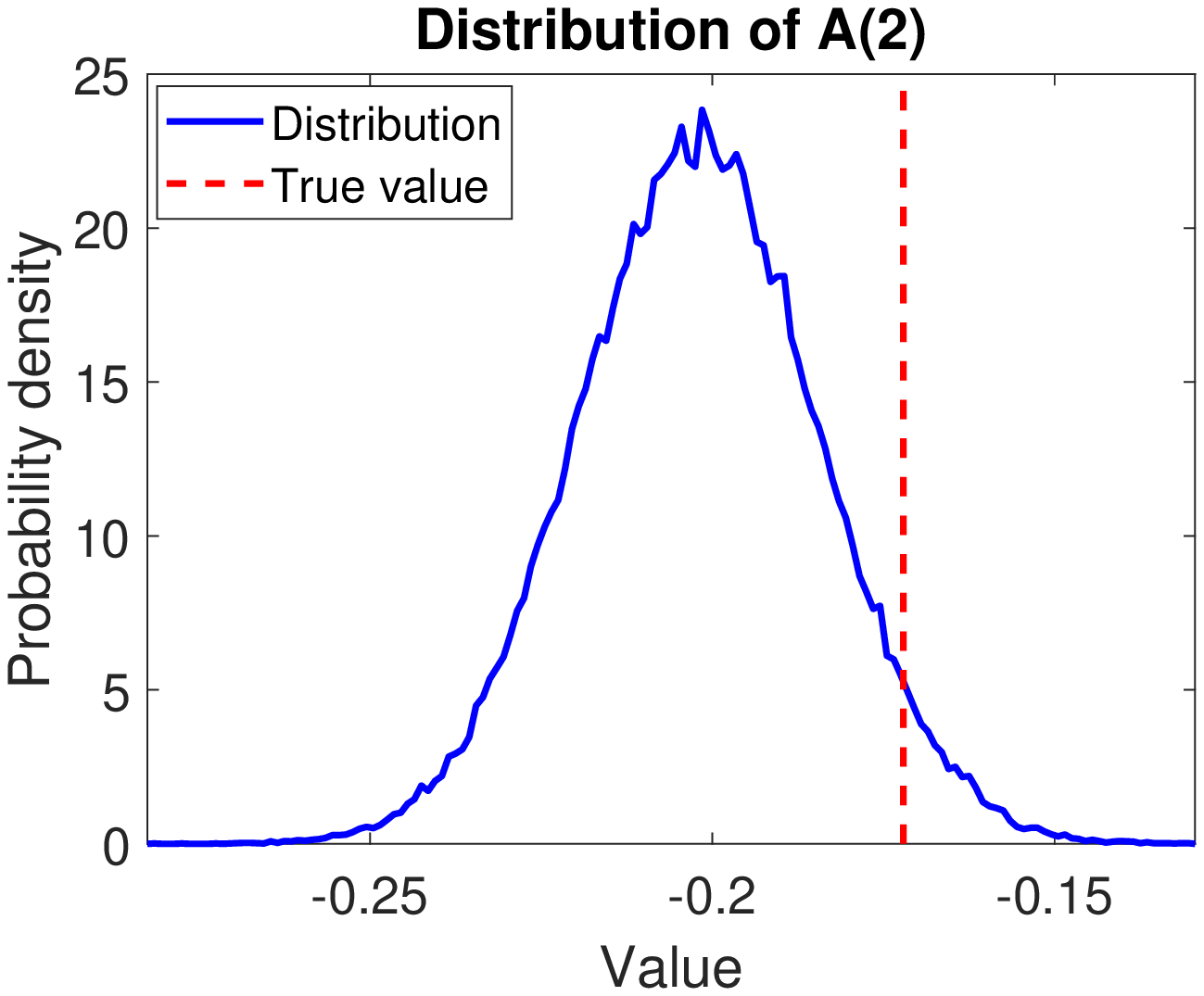}
     }
     \vfill
     \subfloat[Probability distribution of $\bD_1$.]{%
       \includegraphics[width=0.45\textwidth]{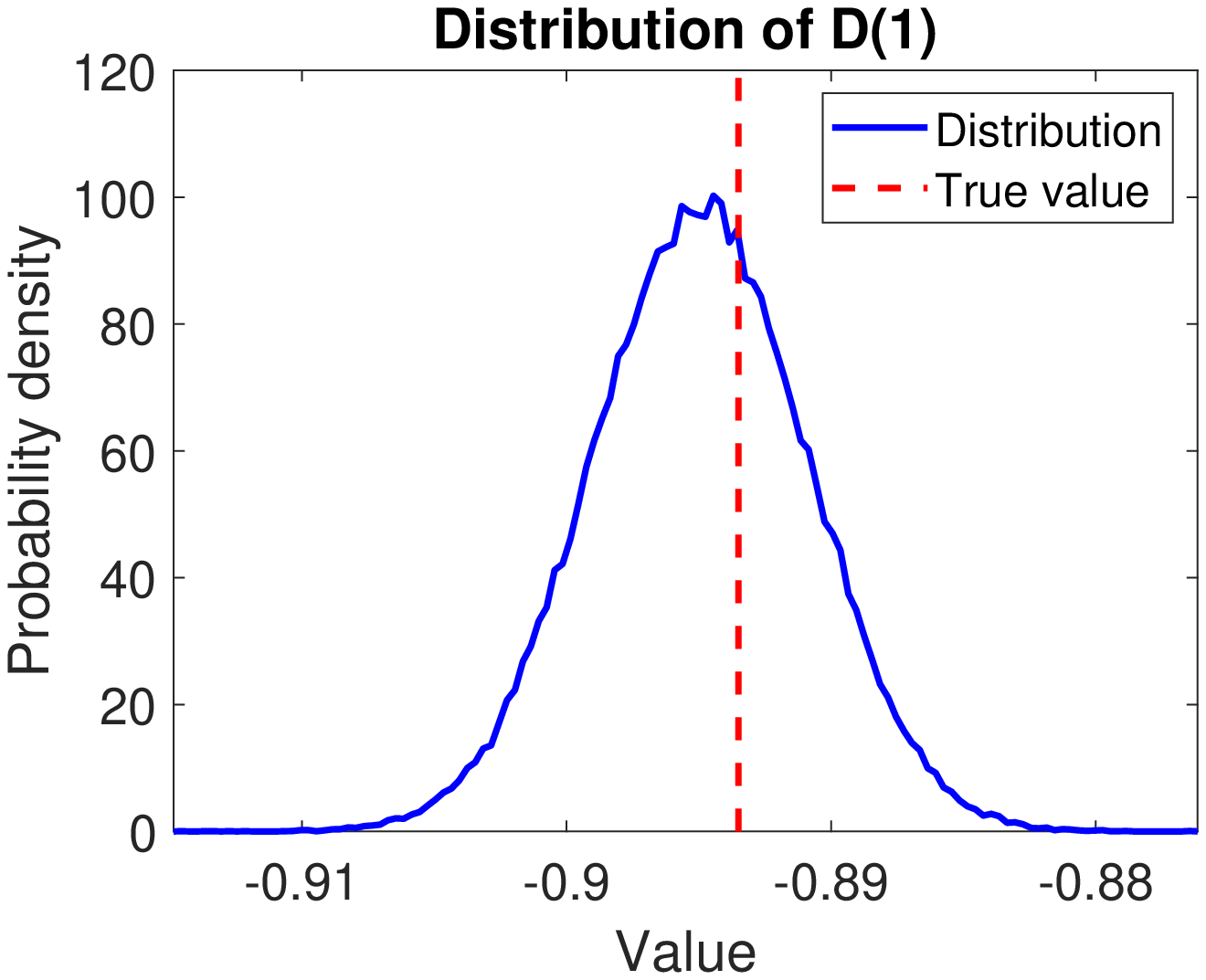}
     }
     \hfill
     \subfloat[Probability distribution of $\bD_2$.]{%
       \includegraphics[width=0.45\textwidth]{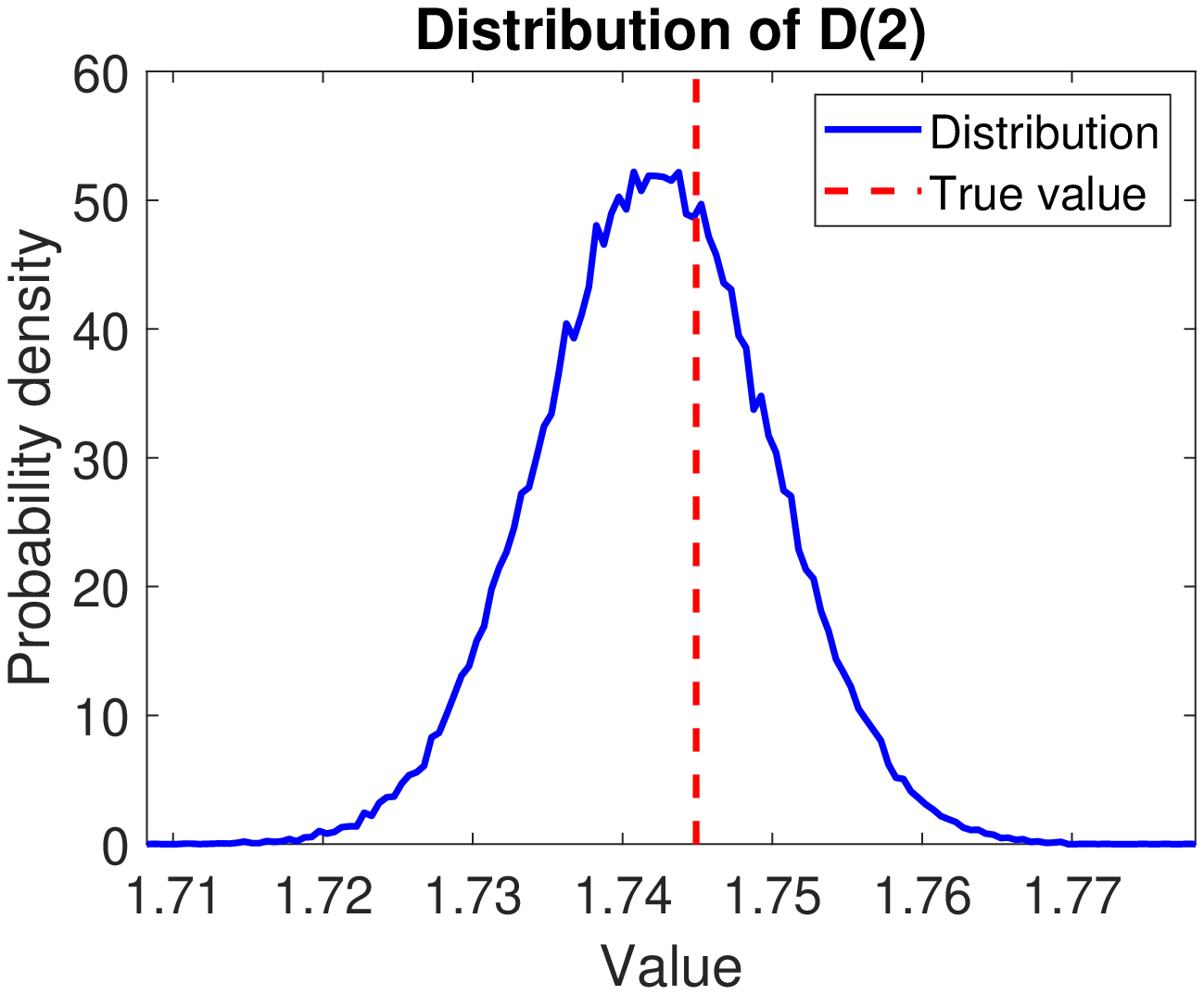}
     }
     \vfill
     \subfloat[Probability distribution of $\bR_1$.]{%
       \includegraphics[width=0.45\textwidth]{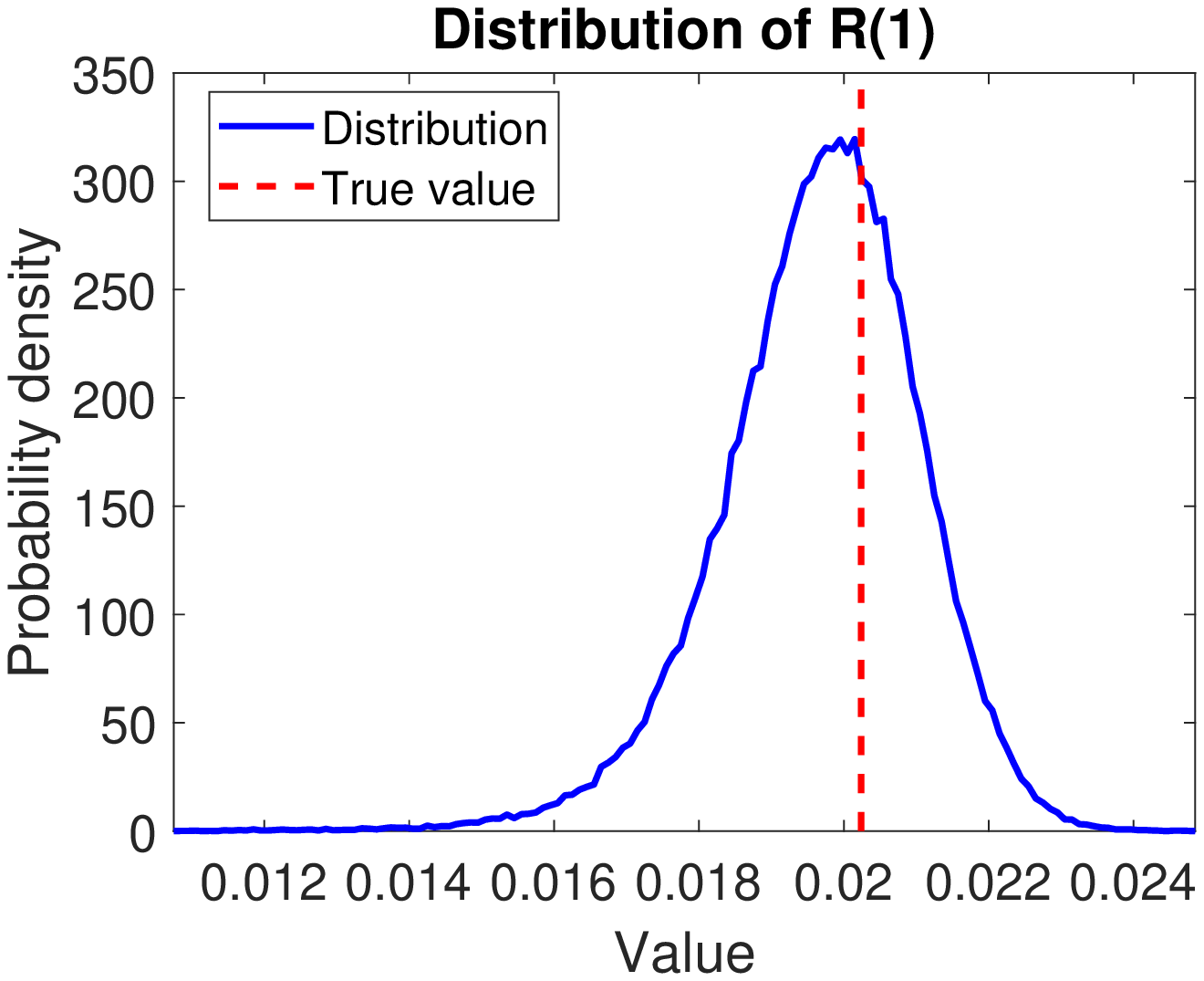}
     }
     \hfill
     \subfloat[Probability distribution of $\bR_2$.]{%
       \includegraphics[width=0.45\textwidth]{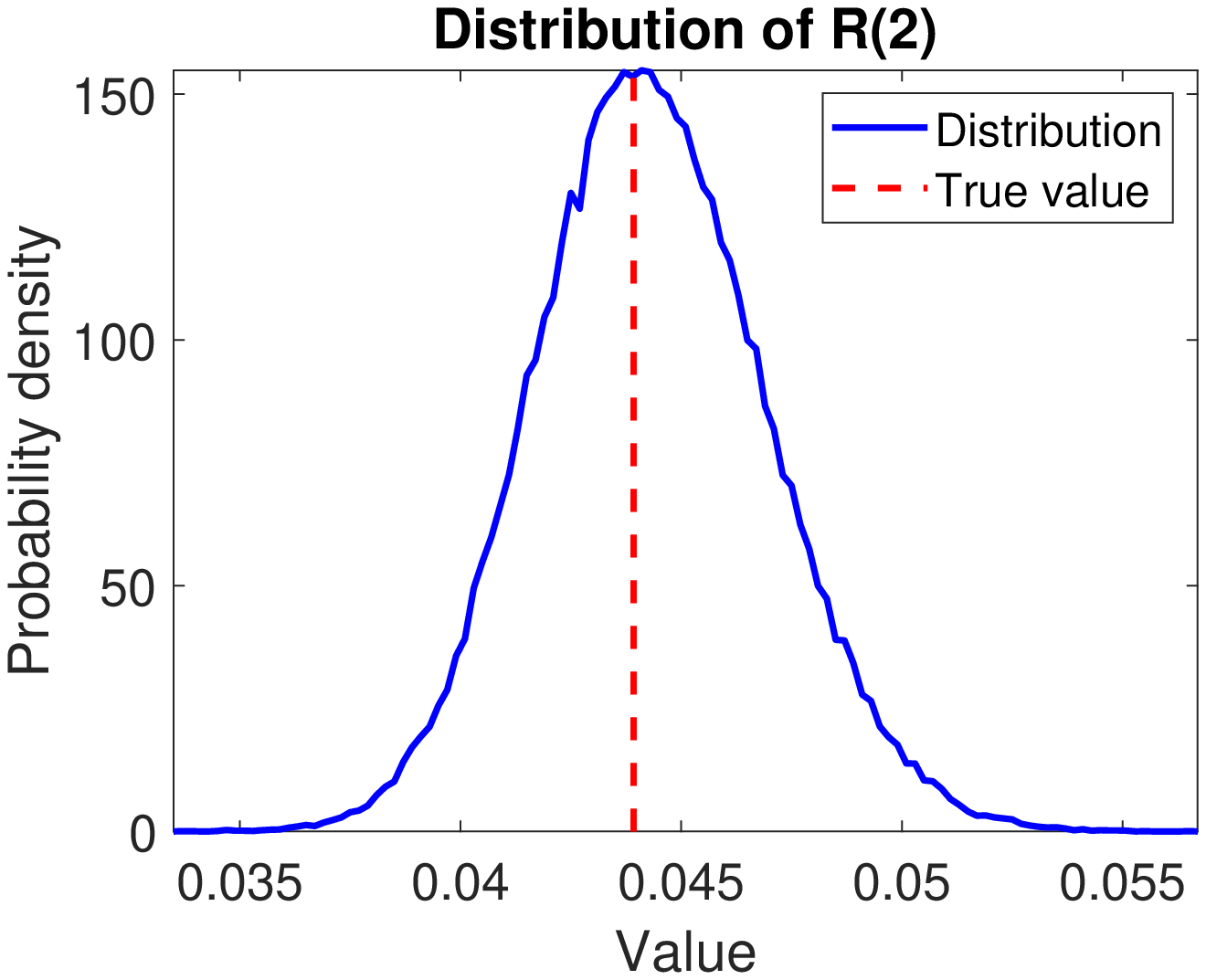}
     }
     \caption{Estimated model parameter distributions for Example 1. The distribution of parameters free from a similarity transformation are shown in solid blue, whereas the true values are indicated by a dashed red vertical line.}
\label{fig:example1res}
\end{figure*}

After $10^5$ iterations of the particle-Gibbs algorithm, the parameter samples $\theta^\ell$ were used to construct Figure~\ref{fig:transitionmatrixprobex1} and Figure~\ref{fig:example1res}.
Figure~\ref{fig:transitionmatrixprobex1} shows the distribution of diagonal elements of the transition matrix, and represents the probability of models being used for consecutive time steps.
As the off-diagonals are constrained by the total law of probability, there is no need to plot them.
Whereas Figure~\ref{fig:example1res} shows the distribution of components within $\{ \bGamma_i,\bPi_i\}_{i=1}^m$ which are free from a similarity transformation.

The proposed method has produced distributions with a large amount of support for the true parameters, which suggests that the algorithm is correctly performing Bayesian identification on the system.
As expected, subsequent testing has shown that an increase in data length will tighten the confidence intervals of the parameters.

Generation of the both figures for this example, required the models to be sorted.
Models were sorted or `relabelled' by comparison of the magnitude of the models Bode response.
This is not a core part of the proposed algorithm, as reordering is only required for plotting purposes. 
\subsection{Multivariate system}
\label{sec:example2}
In this example, we consider identification of a multivariate three-state SISO JMLS system comprising of three models $m=3$.
To the best of the authors knowledge there are no alternative algorithms suitable for this problem, or for generating a ground truth.

As this example considers a multivariate system, there are an infinite number of state space modes which has equivalent system response. 
Because of this, plotting the distribution of the model parameters themselves would be somewhat arbitrary.
Instead, for the analysis, we provide a variation likely frequency responses of the models. 
The distribution of the transition matrix however, is free from similarity transformations, but due to the increase in available models can no longer appear on a 2D plot.

The system analysed in this example, described by the discrete transfer functions and transition matrix
\begin{small}
\begin{subequations}
\begin{align}
H_1(\text{z})&=\frac{217.4 \tz^3 + 212.9 \tz^2 - 0.003827 \tz + 4.603\times 10^{-20}}{ \tz^3 - 1.712 \tz^2 + 0.9512 \tz - 1.481\times 10^{-6}},\\
H_2(\text{z})&=\frac{0.4184 \tz^3 + 0.008764 \tz^2 + 0.1669 \tz - 0.01542}{ \tz^3 - 2.374 \tz^2 + 1.929 \tz - 0.5321},\\
H_3(\text{z})&=\frac{0.2728 \tz^3 - 0.9506 \tz^2 + 1.066 \tz - 0.3881}{\tz^3 - 2.374 \tz^2 + 1.929 \tz - 0.5321 },
\end{align}
and
\begin{align}
\bT &=\begin{bmatrix}0.5 &0.25 &0.25\\
                   0.25 &0.5 &0.25\\
                  0.25 &0.25 &0.5 \end{bmatrix},
\end{align}
\end{subequations}
\end{small}
respectively was then simulated for $N=5000$ time steps using an input $u_k\sim \N(0,1)$.
This system was then used to initialise the proposed procedure to avoid a lengthy burn-in time.
For a real-world example, this initial estimate could be obtained using the EM algorithm \cite{EMJMLSpaperBalenzuela}.

\begin{figure}
     \subfloat[Distribution of the first column of the transition matrix.]{%
       \includegraphics[width=0.45\textwidth]{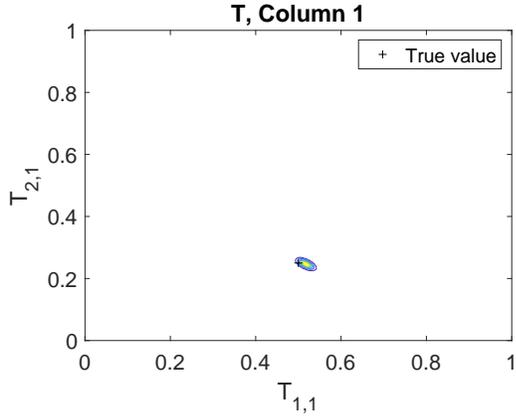}
     }
     \hfill
     \subfloat[Distribution of the second column of the transition matrix.]{%
       \includegraphics[width=0.45\textwidth]{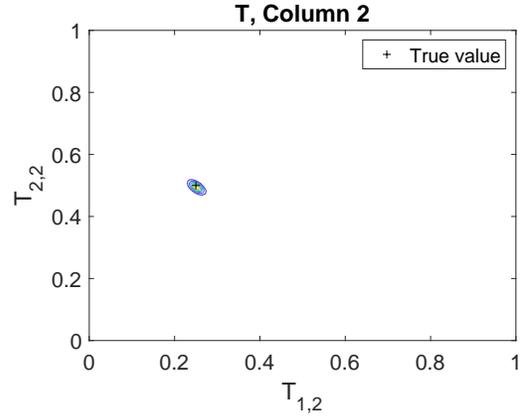}
     }
     \hfill
     \subfloat[Distribution of the third column of the transition matrix.]{%
       \includegraphics[width=0.45\textwidth]{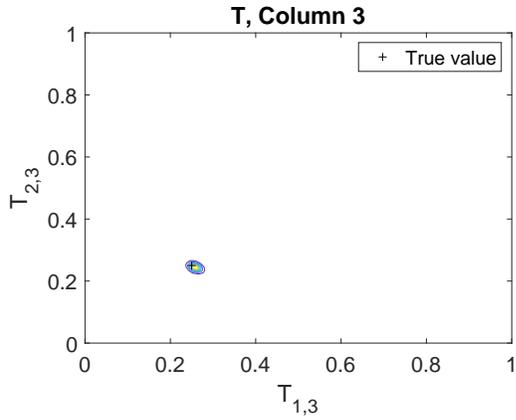}
     }
     \caption{Contour plots showing the estimated transition matrix distributions from Example 2, along with the true values shown with a black +.}
\label{fig:transitionmatrixprobex2}
\end{figure}

\begin{figure}
     \subfloat[Model 1 Bode plot.]{%
       \includegraphics[width=0.4\textwidth]{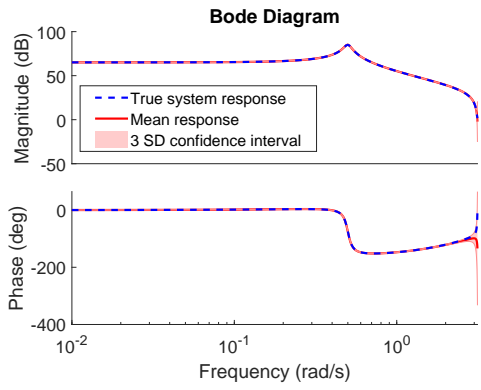}
     }
     \hfill
     \subfloat[Model 2 Bode plot.]{%
       \includegraphics[width=0.4\textwidth]{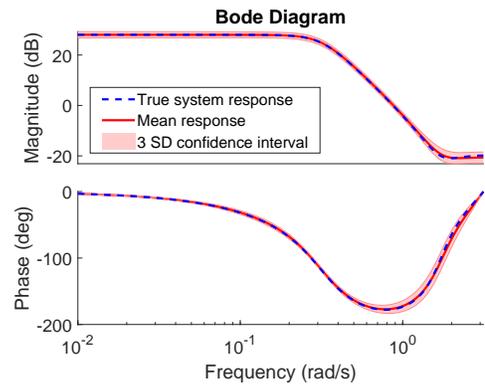}
     }
          \hfill
     \subfloat[Model 3 Bode plot.]{%
       \includegraphics[width=0.4\textwidth]{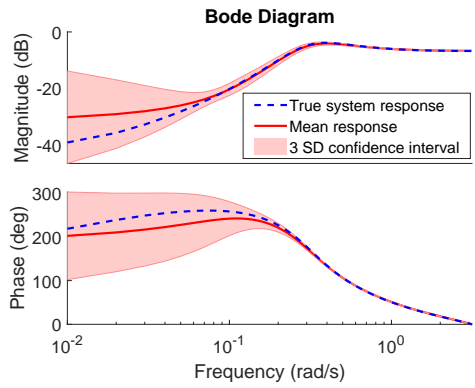}
     }
     \caption{Frequency response from the three models for Example 2. The blue line is the true system response, where the red line and shaded red region represents the estimated mean response and 3 standard deviation confidence region respectively.}
\label{fig:frequencyresponseex2}
\end{figure}

The proposed method was then used to identify the system based on the generated input-output data with the filter being allowed to store $R=5$ hybrid Gaussian mixture components.
The uninformative priors chosen for identification were parameterised by
\begin{subequations} 
\begin{align}
\bM_1=\bM_2=\bM_3=\mathbf{0}_{4\times 4},\\
\bV_1=\bV_2 =\bV_3 = 13\cdot\mathbf{I}_{4\times 4},\\
\bLambda_1=\bLambda_2=\bLambda_3=10^{-10}\cdot\mathbf{I}_{4\times 4},\\
\nu_1=\nu_2=\nu_3=4,\\
\boldsymbol{\alpha} = \mathbf{1}_{3\times 3}.
\end{align}
\end{subequations}
After $10^5$ particle-Gibbs iterations, the samples $\theta^\ell$ were used to construct Figure~\ref{fig:transitionmatrixprobex2} and Figure~\ref{fig:frequencyresponseex2}.
Figure~\ref{fig:transitionmatrixprobex2} shows the estimated distribution of model transition probabilities, and Figure~\ref{fig:frequencyresponseex2} shows the variation of expected model responses.
As with Example 1, models were sorted using their frequency response before producing these figures.
Both of these figures show good support for the model used to generate the data, demonstrating validity of the proposed algorithm. 


\clearpage
\section{Conclusion}\label{sec:conclusion}
We have developed and demonstrated an effective algorithm for Bayesian parameter identification of JMLS systems.
Unlike alternative methods, we have not forced assumptions such as a univariate state or operation according to drift models. 
The proposed method scales easily to an increase in models and state dimension.
This solution was based on a modified version of the discrete particle filter.
In developing this method, points of confusion surrounding the documentation for the DPF were discovered, and have been addressed within this paper.

The proposed method was deployed for Bayesian estimation of a multivariate JMLS system in subsection~\ref{sec:example2}, yielding distributions with good support of the transition matrix and models used to generate the data.
It should be noted that due to a finite data length the true values used are unlikely to align with the maximum \emph{a posteriori} (MAP) estimate.

It is unfortunate that competing approaches for either of the examples don't exist.
Regarding the first example, competing approaches use an inverse-Gamma distribution, which is identical to the inverse-Wishart distribution for the univariate case, and yield an identical algorithm.
Regarding the second example, to the best of our knowledge, no alternative algorithms are available for the unconstrained multivariate JMLS identification problem. 

%
\clearpage
 \bibliographystyle{plain}        
 \bibliography{autosam}           
\appendix
\clearpage

\section{Hybrid state sampling algorithms}
\label{sec:hybridsamplingalgos}
This appendix details a self-contained complete set of algorithms required for sampling hybrid state trajectories.

Here, the forward filter is described by Algorithm~\ref{alg:forward_filter}, that makes use of Algorithm~\ref{alg:DPFResample} to control the computational complexity. Algorithm~\ref{alg:DPFResample} then makes subsequent use of the DPF threshold calculated by Algorithm~\ref{alg:DPFThresh} and systematic resampling scheme detailed in Algorithm~\ref{alg:systematicsampling}.
Following the forward-filtering pass, Algorithm~\ref{alg:backwards_sampler} may be used to sample the required trajectories.

The following notes are intended to assist the practitioner:
\begin{itemize}
  \item The forwards filter and sampling of a latent trajectory is intended to be completed completed using the $\theta^\ell$ parameter set, this is not explicitly written for readability purposes.
\item If the ancestor component has a weight of numerically zero in Algorithm~\ref{alg:forward_filter}, it no longer needs to be tracked, as it cannot be selected by the backwards simulator, improving the produced distribution. 
\item Algorithm~\ref{alg:DPFThresh} should be implemented using the LSE trick.
\item Duplicated components are allowed to be returned from Algorithm~\ref{alg:systematicsampling}.
\item When implementing Algorithm~\ref{alg:backwards_sampler}, it is beneficial (but not required) to use the non-reduced forwards filtered distribution from Algorithm~\ref{alg:forward_filter}.
\end{itemize}

Before filtering and backwards simulating each time step using Algorithms \ref{alg:forward_filter} and \ref{alg:backwards_sampler}, the following transformation made to efficiently handle the cross-covariance term $\bS_{z_k}$, 
\begin{align}
&\bar{\mathbf{A}}_{z_k} = \mathbf{A}_{z_k}-\mathbf{S}_{z_k}\mathbf{R}_{z_k}^{-1} \mathbf{C}_{z_k}, \nonumber \\
&\bar{\mathbf{B}}_{z_k} = \begin{bmatrix}\mathbf{B}_{z_k} -\mathbf{S}_{z_k}\mathbf{R}_{z_k}^{-1}\mathbf{D}_{z_k} &\ \mathbf{S}_{z_k}\mathbf{R}_{z_k}^{-1} \end{bmatrix},\nonumber \\
&\bar{\mathbf{C}}_{z_k} =\mathbf{C}_{z_k},\quad \bar{\mathbf{D}}_{z_k} = \begin{bmatrix} \mathbf{D}_{z_k} & \ \mathbf{0}_{n_y} \end{bmatrix},\nonumber \\
&\bar{\mathbf{Q}}_{z_k}=\mathbf{Q}_{z_k}  -\mathbf{S}_{z_k}\mathbf{R}_{z_k}^{-1}\mathbf{S}^T_{z_k},\quad \bar{\mathbf{R}}_{z_k}=\mathbf{R}_{z_k}, \nonumber \\
&\bar{u}_k = \begin{bmatrix}u_k\\y_k\end{bmatrix}.
\label{eq:removeStransformation}
\end{align}
Notice that the new system uses a different input $\bar{u}_k$. 
\begin{algorithm}
  \begin{small}
\caption{Forward Filter}
\label{alg:forward_filter}
\begin{algorithmic}[1]
  \Require Prior joint distribution $p(x_1,z_1)$ given by
  \begin{align}
    \label{eq:MCHA6100 Notes on JMLS:7}
    p(x_1,z_1) = \sum_{i=1}^{M^p_1(z_1)} w^{i}_{1|0}(z_1) 
    \mathcal{N}\left(x_1|\mu^{i}_{1 | 0}(z_1),  \bP^{i}_{1 | 0}(z_1)\right),
  \end{align}
  where $M^p_1(z_1)=1.$
  maximum number of hybrid components $M$ and model parameters $\theta$.
  \State Set the conditioned ancestor particle index number $a_1= 1$.
  Note that if this is the first particle-Gibbs iteration i.e. $\ell=1$, filtering does not need to be conditioned on a model sequence, alternatively it can be set with any valid random sequence satisfying $z_k^{\ell-1} \in \{1,\dots,m\}\,  \forall k=1,\dots,N+1$ .
  \For{$k=1$ to $N$}
  \For{$z_k=1$ to $m$}
  \For{$i=1$ to $M_k^p(z_k)$}
\begin{subequations}
\label{eq:fwd_fltr}
\label{eq:JMLSFcorrect}
\begin{align}
    \tilde{w}^{i}_{k | k}(z_k) &= w^{i}_{k | k-1}(z_k) \mathcal{N}\left (y_k|
                               \eta_{k|k}^{i}(z_k),
                               \bXi_{k|k}^{i}(z_k)\right
                               ),\label{eq:filtlikelihood43}\\
\mu^{i}_{k | k}(z_k) &= \mu^{i}_{k | k-1}(z_k) + \bK_{k|k}^{i}(z_k)[y_k - \eta_{k|k}^{i}(z_k)],\\
    \eta_{k|k}^{i}(z_k) &= \bar{\bC}_{z_k} \mu^{i}_{k | k-1}(z_k) + \bar{\mathbf{D}}_{z_k}\bar{u}_k,\\
    \bP^{i}_{k | k} (z_k) &=  \bP^{i}_{k | k-1}(z_k)  - \bK_{k|k}^{i}(z_k) \bar{\bC}_{z_k} \bP^{i}_{k | k-1}(z_k),\label{eq:Pfilt43}\\ 
\bK_{k|k}^{i}(z_k) &= \bP^{i}_{k | k-1} (z_k) \bar{\bC}_{z_k}^T (\bXi_{k|k}^{i}(z_k))^{-1},\\   
    \bXi_{k|k}^{i}(z_k) &= \ \bar{\bC}_{z_k} \bP^{i}_{k | k-1}(z_k) \bar{\bC}_{z_k}^T+ \bar{\bR}_{z_k}. \label{eq:sigmafilt43}
  \end{align} 
  %

\end{subequations}
\EndFor (over $i$) 
\EndFor (over $z_k$) 
\begin{align}
  \label{eq:12} 
   & w^{i}_{k|k}(z_k) = \frac{\tilde{w}^{i}_{k | k}(z_k) }{\sum_{z_k=1}^m  \sum_{i=1}^{M_k^p(z_k)} \tilde{w}^{i}_{k | k}(z_k)}
    \nonumber \\& \forall z_k=1,\dots,m \text{ and } \forall i=1,\dots,M^p_k(z_k) 
\end{align} 
\If{$\sum_{z=1}^m M_k^p(z)> M$}
\State Using Algorithm~\ref{alg:DPFResample} resample the hybrid mixture defined by $\{w_{k|k}(z_k),\mu_{k|k}(z_k),\bP_{k|k}(z_k)  \}_{i=1}^{M^p_k(z_k)} \forall z_k$ to yield a replacement mixture, preserving the conditioned component $\{w_{k|k}^{a_{k}}(z_k^{\ell-1}),\mu_{k|k}^{a_{k}}(z_k^{\ell-1}),\bP_{k|k}^{a_{k}}(z_k^{\ell-1}) \}$.
\EndIf
\For{$z_{k+1}=1$ to $m$}
\State Set $j=0$.
\For{$z_k=1$ to $m$}
\For{$i=1$ to $M^p_{k}(z_k)$}
\State Increment $j$.
\If{$i=a_k$ and $z_k=z_k^{\ell-1}$}
\State Set $a_{k+1} = j$.
\EndIf
\begin{subequations}
\begin{align}
  {w}^{j}_{k+1 | k}(z_{k+1}) &=  \mathbf{T}_{z_{k+1} , z_k}  w_{k|k}^{i}(z_k)  ,\\ 
  \mu^{j}_{k+1 | k}(z_{k+1}) &= \bar{\bA}_{z_k} \mu_{k | k}^{i}(z_k) + \bar{\bB}_{z_k}\bar{u}_k ,\\
  \bP^{j}_{k+1 | k}(z_{k+1}) &= \bar{\bA}_{z_k} \bP^{i}_{k | k}(z_k) \bar{\bA}_{z_k}^T + \bar{\bQ}_{z_k}. \label{eq:predPfilt43}
\end{align}
\end{subequations}
\EndFor (over $i$)
\EndFor (over $z_k$)
\EndFor (over $z_{k+1}$)
\EndFor (over $k$)
\end{algorithmic}
  \end{small}
\end{algorithm}

\begin{algorithm}
  \begin{small}
\caption{DPF Resampling}
\label{alg:DPFResample}
\begin{algorithmic}[1]
  \Require The maximum number of components which may be kept $M$, and a hybrid mixture which needs to be reduced that contains an ancestor component.
  \State Place the ancestor component into set $S$ and all other components into set $\bar{S}$.
   \State Form normalised weights for the $\bar{S}$ set
  \begin{align} \alphaaa^i = \frac{w^i}{\sum_{i \in \bar{S}} w^i} \quad \forall i \in \bar{S} \end{align}
  \State Call Algorithm~\ref{alg:DPFThresh} with $\{\alphaaa^i\}_{i \in \bar{S}}$ to discern the number of components to keep deterministically $L$, with a maximum number of stored components being $K=M-1$.
  \State Place $L$ components with the highest $\alphaaa^i$ values into set $S$ with their original weights $w^i$, removing them from set $\bar{S}$.
  \State Renormalise the new set $\bar{S}$,  
  \begin{align} \alphaaa^i = \frac{w^i}{v} \quad \forall i \in \bar{S}, \text{ where } v=\sum_{i \in \bar{S}} w^i. \end{align}
  \State Using Algorithm~\ref{alg:systematicsampling} perform systematic sampling \cite{hol2004resampling} to sample $R=M-L-1$ components from set $\bar{S}$ based upon weights $\{\alphaaa^i\}_{i \in \bar{S}}$, placing them in set $S$ with a new weight of \begin{align} w^i =\frac{v}{R}.\end{align}
  \State Return the reduced mixture governed by the set $S$.
\end{algorithmic}
  \end{small}
\end{algorithm}

\begin{algorithm}
  \begin{small}
\caption{Compute DPF Threshold}
\label{alg:DPFThresh}
\begin{algorithmic}[1]
  \Require Maximum number of components which may be kept $K$, and mixture weights $\alphaaa^i \, \forall i=1,\dots,n$.
  \State Sort weights in descending order $\{\alphaaa^i\}_{i=1}^n$, e.g. $\alphaaa_1 \geq \alphaaa_2$.
  \State Set $L=0$.
  \For{$j=1$ to $K$}
    \If{$\alphaaa^j(K - j) \geq \sum_{i=j+1}^n \alphaaa^i$}
    \State Set $L = j$.
    \Else
    \State Return number of components to keep $L$.
    \EndIf
  \EndFor
  \State Return number of components to keep $L$.
\end{algorithmic}
  \end{small}
\end{algorithm}

\begin{algorithm}
  \begin{small}
\caption{Systematic Sampling}
\label{alg:systematicsampling}
\begin{algorithmic}[1]
  \Require Number of components to be sampled $R$, and mixture weights $\alphaaa^i \, \forall i=1,\dots,n$.
  \State Sample from a uniform distribution $u \sim \mathcal{U}$(0,1).
  \State Compute cumulative probability mass function, such that $Q(i) = \sum_{j=1}^i \alphaaa^j \, \forall i=1,\dots,n$.
  \For{$j=1$ to $R$}
    \For{$i=1$ to $n$}
      \If{$Q(i) \geq (j-1+u)/R$}
        \State Keep component $i$.
        \State Break.
      \EndIf
    \EndFor
  \EndFor
\end{algorithmic}
  \end{small}
\end{algorithm}

\begin{algorithm}
  \begin{small}
\caption{Backwards Sampler}
\label{alg:backwards_sampler}
\begin{algorithmic}[1]
  \Require A Forwards filtered distribution from Algorithm~\ref{alg:forward_filter}.
  \State Sample $z_{N+1}^\ell$, $b_{N+1}$ and $x_{N+1}^\ell$
  \begin{align}
    \label{eq:9}
    z_{N+1}^\ell &\sim \mathcal{C} \left ( \left \{ \sum_{i=1}^{M^p_{N+1}(j)} w^{i}_{N+1 | N}(j) \right
               \}_{j=1}^{m} \right ),\\
    b_{N+1} &\sim \mathcal{C} \left ( \left \{ \frac{w^{i}_{N+1 | N}(z_{N+1}^\ell) }{ \sum_{j=1}^{M_{N+1}^p(z_{N+1}^\ell)} w^{j}_{N+1 | N}(z_{N+1}^\ell)} \right
              \}_{i=1}^{M_{N+1}^p(z_{N+1}^\ell)} \right ),\\
    x_{N+1}^\ell &\sim \mathcal{N} \left ({x_{N+1}} |
                   \mu^{b_{N+1}}_{N+1|N}(z_{N+1}^\ell),\, \bP^{b_{N+1}}_{N+1|N} (z_{N+1}^\ell)\right )
  \end{align}
  \For{$k=N$ to $1$}
    \For{$z_k=1$ to $m$}
  \For{$i=1$ to $M^p_k(z_k)$}
\begin{subequations}
  \begin{align}
    \tilde{w}^{i}_{k | N}(z_k) &= \mathbf{T}_{z_{k+1}^\ell,z_k} 
                                w_{k|k}^{i} (z_k) \mathcal{N} \left (x_{k+1}^\ell| \eta^{i}_{k|N}(z_k)
                                                                ,
                                                                \bXi_{k|N}^{i}(z_k)
                                                                \right) \label{eq:likesmooth34},\\
   \mu^{i}_{k | N}(z_k) &= \mu^{i}_{k|k}(z_k) + \bK_{k|N}^{i}(z_k) [x_{k+1}^\ell -\eta^{i}_{k|N}(z_k)],\\
    \eta_{k|N}^{i}(z_k) &= \bar\bA_{z_k}\mu^{i}_{k|k}(z_k)+\bar\bB_{z_k}\bar{u}_k,\\
    \bP^{i}_{k | N}(z_k) &= \left(\bI- \bK_{k|N}^{i}(z_k) \bar\bA_{z_k} \right)\bP^{i}_{k|k}(z_k), \label{eq:Psmooth34} \\
\bK_{k|N}^{i}(z_k) &= \bP^{i}_{k|k}(z_k) \bar\bA^T_{z_k} (\boldsymbol{\Xi}_{k|N}^{i}(z_k))^{-1},\\
\boldsymbol{\Xi}_{k|N}^{i}(z_k) &= \bar\bA_{z_k} \bP^{i}_{k|k}(z_k) \bar\bA^T_{z_k} + \bar\bQ_{z_k} \label{eq:xismooth34},
  \end{align}
\end{subequations}
\EndFor (over $i$)
\EndFor (over $z_k$)
\begin{align}
{w}_{k|N}^{i} (z_k)= \frac{\tilde{w}_{k|N}^{i}(z_k)}{\sum_{z_k=1}^m   \sum_{i=1}^{M^p_k(z_k)}  \tilde{w}_{k|N}^{i}(z_k)} \nonumber \\
\forall z_k=1,\dots,m, \text{ and } \forall i=1,\dots,M^p_k(z_k) .
\end{align}
  \State Sample $z_{k}^\ell$, $b_{k}$ and $x_{k}^\ell$
  \begin{align}
    \label{eq:10}
    z_{k}^\ell &\sim \mathcal{C} \left ( \left \{ \sum_{i=1}^{M^p_k(j)} w^{i}_{k | N}(j) \right
               \}_{j=1}^{m} \right ),\\
    b_{k} &\sim \mathcal{C} \left ( \left \{ \frac{w^{i}_{k| N}(z_{k}^\ell) }{ \sum_{j=1}^{M^p_k(z_k^\ell)} w^{j}_{k | N} (z_{k}^\ell)} \right
              \}_{i=1}^{M^p_k(z_k^\ell)} \right ),\\
    x_{k}^\ell &\sim \mathcal{N} \left (x_{k} |
                   \mu^{b_{k}}_{k|N}(z_{k}^\ell),\, \bP^{b_{k}}_{k|N}(z_{k}^\ell)\right )
  \end{align}
\EndFor (over $k$)
\end{algorithmic}
\end{small}
\end{algorithm}
\newpage
\section{Proofs}
In this appendix we provide proofs for the Algorithms and Lemmata within the paper.

\subsection{Proof of Algorithm~\ref{alg:backwards_sampler}}
\begin{small}
For readability, in this proof we utilise the shorthand $\theta^\ell = \left\{ \{\bGamma^\ell_i, \bPi^\ell_i \}_{i=1}^m, \bT^\ell \right\}$.
We begin the derivation with
\begin{align}
&p(x_{1:N+1},z_{1:N+1}|\{\bGamma^\ell_i, \bPi^\ell_i \}_{i=1}^m, \bT^\ell,y_{1:N})  = p(x_{N+1},z_{N+1}|\theta^\ell,y_{1:N})  \prod_{k=1}^N \frac{p(x_{k+1},z_{k+1}|x_{k},z_{k},\theta^\ell,y_{1:k}) p(x_{k},z_{k}|\theta^\ell,y_{1:k})}{p(x_{k+1},z_{k+1}|\theta^\ell,y_{1:k})} .
\end{align}
%
%
%
We will now outline computations the distribution $\frac{p(x_{k+1}^\ell,z_{k+1}^\ell|x_{k},z_{k},\theta^\ell,y_{1:k}) p(x_{k},z_{k}|\theta^\ell,y_{1:k})}{p(x_{k+1}^\ell,z_{k+1}^\ell|\theta^\ell,y_{1:k})}$.
Since the terms in the numerator have a known form,
\begin{align}
&p(x_{k},z_{k}|\theta^\ell,y_{1:k}) = \sum_{i=1}^{M_k^f(z_k)} w_{k|k}^i(z_k)\N(x_k|\mu^i_{k|k}(z_k),\mathbf{P}^i_{k|k}(z_k)),
\end{align}
and by removing conditionally independent terms we yield
\begin{align}
&p(x_{k+1}^\ell,z_{k+1}^\ell|x_{k},z_{k},\theta^\ell,y_{1:k})   =\Prob(z_{k+1}^\ell| z_{k},\theta^\ell)p(x_{k+1}^\ell|x_{k},z_{k},\theta^\ell,y_{1:k})   =  \mathbf{T}^\ell_{z_{k+1}^\ell,z_k} \N(x_{k+1}^\ell|\bar\bA^\ell_{z_k}x_k+\bar\bB^\ell_{z_k}\bar{u}_k,\bar\bQ^\ell_{z_k}),
\end{align}
we can rewrite and manipulate the numerator as follows,
\begin{align}
&p(x_{k},z_{k}|\theta^\ell,y_{1:k})p(x_{k+1}^\ell,z_{k+1}^\ell|x_{k},z_{k},\theta^\ell,y_{1:k})  \nonumber \\
&= \sum_{i=1}^{M_k^f(z_k)} w_{k|k}^i(z_k)\N(x_k|\mu^i_{k|k}(z_k),\bP^i_{k|k}(z_k))\mathbf{T}^\ell_{z_{k+1}^\ell,z_k}  \N(x_{k+1}^\ell|\bar\bA^\ell_{z_k}x_k+\bar\bB^\ell_{z_k}\bar{u}_k,\bar\bQ^\ell_{z_k}) \nonumber \\
&= \sum_{i=1}^{M_k^f(z_k)}  \tilde{w}_{k|N}^i(z_k) \N(x_k|\mu^i_{k|N}(z_k),{\bP}^i_{k|N}(z_k)),
\end{align}
where $\tilde{w}_{k|N}^i(z_k)$, and $\mu^i_{k|N}$, and ${\bP}^i_{k|N}$ can be computed straight-forwardly as this pattern of Normal distribution terms has a well known solution which is identical the correction step of the weighted Kalman filter used for forward-filtering.
With the numerator of $\frac{p(x_{k+1}^\ell,z_{k+1}^\ell|x_{k},z_{k},\theta^\ell,y_{1:k}) p(x_{k},z_{k}|\theta^\ell,y_{1:k})}{p(x_{k+1}^\ell,z_{k+1}^\ell|\theta^\ell,y_{1:k})}$ now having a closed form, it can be marginalised to yield the denominator
\begin{align}
p(x_{k+1}^\ell,z_{k+1}^\ell|\theta^\ell,y_{1:k}) 
&= \sum_{z_k=1}^m   \sum_{i=1}^{M_k^f(z_k)}  \tilde{w}_{k|N}^i(z_k).
\end{align}
Therefore 
\begin{align}
&p(x_{k},z_{k}|x_{k+1:N+1},z_{k+1:N+1},\theta^\ell,y_{1:N})  = \sum_{i=1}^{M_k^f(z_k)}  {w}_{k|N}^i(z_k) \N(x_k|\mu^i_{k|N}(z_k),{\bP}^i_{k|N}(z_k)),
\end{align}
where 
\begin{align}
{w}_{k|N}^i(z_k) = \frac{\tilde{w}_{k|N}^i(z_k)}{\sum_{z_k=1}^m   \sum_{i=1}^{M_k^f(z_k)}  \tilde{w}_{k|N}^i(z_k)}.
\end{align}

Sampling may be computed by introducing a auxiliary variable $b$ of some hybrid Gaussian mixture, which represents a possible model sequence for this application,
\begin{align}
\label{eq:conditx438765}
p(x,z,b|\cdot) &= w^b(z)\N(x|\mu^b(z),\bP^b(z)).
\end{align}
Normally this variable is not of interest and is marginalised out. 
Using conditional probability, we outline sampling from $x,z$, and $b$ as sampling from the distributions
\begin{align}
p(x,z,b|\cdot) = p(x|z,b,\cdot)\Prob(b|z,\cdot)\Prob(z|\cdot),
\end{align}
where the $b$ component of the sample may be discarded to obtain a sample from $p(x,z|\cdot)$.
Next we consider marginalisation of $x$ to yield
\begin{align}
\label{eq:new454565}
\Prob(z,b|\cdot) &=\int p(x,z,b|\cdot) \, dx  =\int  w^b(z)\N(x|\mu^b(z),\bP^b(z)) \, dx = w^b(z).
\end{align}
Considering the $\Prob(b|z,\cdot)$ and $\Prob(z|\cdot)$ distributions,
\begin{align}
p(z|\cdot) &= \sum_{b=1}^{M(z)} \Prob(z,b|\cdot) = \sum_{b=1}^{M(z)} w^b(z), \\
p(b|z,\cdot) &= \frac{\Prob(z,b|\cdot)}{\Prob(z|\cdot)} = \frac{w^b(z)}{\sum_{b=1}^{M(z)} w^b(z)},
\end{align}
where these are Categorical distributions, which $z^\ell$ and $b$ can easily be sampled from.
Next it follows from \eqref{eq:conditx438765} and \eqref{eq:new454565} that
\begin{align}
&p(x|z,b,\cdot) = \frac{p(x,z,b|\cdot)}{\Prob(z,b|\cdot)}=\frac{w^b(z)\N(x|\mu^b(z),\bP^b(z))}{w^b(z)}  =\N(x|\mu^b(z),\bP^b(z)),
\end{align}
and therefore sampling $x^\ell$ can be completed straight-forwardly using
\begin{align}
x^\ell \sim \N(x|\mu^{b}(z^\ell),\bP^{b}(z^\ell)) .
\end{align}
\qed
\end{small}

\subsection*{Proof of Lemma~\ref{lem:correctedconjucate3332}}
\begin{small}
For readability, in this proof we utilise the shorthand $\theta = \left\{ \{\bGamma_i, \bPi_i \}_{i=1}^m, \bT \right\}$.
We begin with
\begin{align}
&p(\theta|x^\ell_{1:N+1},z^\ell_{1:N+1},y_{1:N})  
\propto p(\theta) \prod_{k=1}^N \Prob(z_{k+1}^\ell|x^\ell_{k+1},z^\ell_k,x^\ell_k,y_{1:k}, \theta)  p(x^\ell_{k+1},y_{k}|x^\ell_k,z^\ell_k,y_{1:k-1}, \theta)  p(x^\ell_{k+1},y_{k}|x^\ell_k,z^\ell_k,y_{1:k-1}, \theta) ,
\end{align}
as $p(x^\ell_1,z^\ell_1)$ is a constant for each iteration.
By expanding $\theta$ and exercising conditional independence, we yield
\begin{align}
&p(\theta|x^\ell_{1:N+1},z^\ell_{1:N+1},y_{1:N})  
\propto
(p(\bT) \prod_{k=1}^N \Prob(z_{k+1}^\ell|z^\ell_k,\bT))    (\prod^m_{i=1}p(\bGamma_i|\bPi_i)p(\bPi_i))     \prod_{k=1}^N p(x^\ell_{k+1},y_{k}|x^\ell_k, \bGamma_{z^\ell_k}, \bPi_{z^\ell_k}  ).
\end{align}
Next we consider the each of the columns in the $\bT$ matrix, written as $\bT_i$, to be conditionally independent, and therefore $p(\bT) = \prod_{i=1}^m p(\bT_i)$. Therefore
\begin{align}
&p(\theta|x^\ell_{1:N+1},z^\ell_{1:N+1},y_{1:N})  
\propto
(  \prod_{i=1}^m p(\bT_i) \prod_{k \in G_i} \Prob(z_{k+1}^\ell|\bT_i))      (\prod^m_{i=1}p(\bGamma_i|\bPi_i)p(\bPi_i)   \prod_{k\in G_i} p(x^\ell_{k+1},y_{k}|x^\ell_k, \bGamma_i, \bPi_i  ) ), \nonumber \\
\end{align}
where $G_i$ is the set of time steps in sample $\ell$ which has model $i$ being active, i.e. $i = z^\ell_k, k\in G_i$.
Substituting the assumed distributions for these terms yields
\begin{align}
p(\theta|x^\ell_{1:N+1},z^\ell_{1:N+1},y_{1:N})  
\propto &(  \prod_{i=1}^m \D(\bT_i|\boldsymbol\alpha_i) \prod_{k \in G_i} \C(z^\ell_{k+1}|\bT_i) )     (\prod^m_{i=1}  \MN(\bGamma_i |\bM_i,\bPi_i,\bV_i)\IW(\bPi_i| \bLambda_i, \nu_i) \nonumber \\
&\quad\cdot    \prod_{k\in G_i} 
\N(\begin{bmatrix} y_{k} \\ x^\ell_{k+1}\end{bmatrix} \bigg| \bGamma_i\begin{bmatrix}x_k^\ell \\ u_k \end{bmatrix}, \bPi_i) ) ,
\end{align}
where $\boldsymbol{\alpha}_i$ denotes the $i$-th column of $\boldsymbol{\alpha}$.
We can now use well known results for updating the conjugate prior, see e.g. \cite{wills2012estimation}.
This yields a solution of the form
\begin{align}
\label{eq:finalsoln765432}
p&(\{\bGamma_i, \bPi_i \}_{i=1}^m, \bT|x^\ell_{1:N+1},z^\ell_{1:N+1},y_{1:N})  \propto (  \prod_{i=1}^m \D(\bT_i|\bar{\boldsymbol\alpha}_i)  )    (\prod^m_{i=1}  \MN(\bGamma_i |\bar\bM_i,\bPi_i,\bar\bV_i)\IW(\bPi_i| \bar\bLambda_i, \bar\nu_i)  ),
\end{align}
where the parameters can be calculated using the instructions provided in Lemma~\ref{lem:correctedconjucate3332}.
A sample from this distribution can be taken by sampling from the Dirichlet, Matrix-Normal and Inverse-Wishart distribution for each model.
\qed
\end{small}
\end{document}

%% file: usr_macros.tex
\usepackage{stackengine}
\def\define{\mathrel{\ensurestackMath{\stackon[1pt]{=}{\scriptstyle\Delta}}}} 
\newcommand{\real}[0]{\mathbb{R}} 
\newcommand{\Prob}[0]{\mathbb{P}} 
\newcommand{\Tr}[0]{\text{\normalfont tr}} 
\newcommand{\vect}[0]{\text{\normalfont Vec}} 
\newcommand{\N}[0]{\mathcal{N}} 
\newcommand{\MN}[0]{\mathcal{M}\mathcal{N}} 
\newcommand{\IW}[0]{\mathcal{W}^{-1}} 
\newcommand{\D}[0]{\mathcal{D}} 
\newcommand{\C}[0]{\mathcal{C}} 
\newcommand{\bA}[0]{\mathbf{A}} 
\newcommand{\bB}[0]{\mathbf{B}} 
\newcommand{\bC}[0]{\mathbf{C}} 
\newcommand{\bD}[0]{\mathbf{D}} 
\newcommand{\bH}[0]{\mathbf{H}} 
\newcommand{\bT}[0]{\mathbf{T}} 
\newcommand{\bM}[0]{\mathbf{M}} 
\newcommand{\bK}[0]{\mathbf{K}} 
\newcommand{\bI}[0]{\mathbf{I}} 
\newcommand{\bP}[0]{\mathbf{P}} 
\newcommand{\bV}[0]{\mathbf{V}} 
\newcommand{\bQ}[0]{\mathbf{Q}} 
\newcommand{\bR}[0]{\mathbf{R}} 
\newcommand{\bS}[0]{\mathbf{S}} 
\newcommand{\balpha}[0]{\boldsymbol{\alpha}} 

\newcommand{\bGamma}[0]{\boldsymbol{\Gamma}} 
\newcommand{\bPi}[0]{\boldsymbol{\Pi}} 
\newcommand{\bLambda}[0]{\boldsymbol{\Lambda}} 
\newcommand{\bPhi}[0]{\boldsymbol{\Phi}} 
\newcommand{\bSigma}[0]{\boldsymbol{\Sigma}} 
\newcommand{\bPsi}[0]{\boldsymbol{\Psi}} 
\newcommand{\bXi}[0]{\boldsymbol{\Xi}} 

\newcommand{\tz}{\text{z}}